\documentclass[]{article}

\usepackage[affil-it]{authblk}
\usepackage{a4wide}
\usepackage{color}
\usepackage[normalem]{ulem}
\usepackage[ruled,vlined]{algorithm2e}
\usepackage{hyperref}
\usepackage{amssymb}
\usepackage{graphicx}
\usepackage{amsfonts}
\usepackage{multirow} 
\usepackage{booktabs}
\usepackage[english]{babel}
\usepackage[sort&compress,square,comma,numbers]{natbib}

\providecommand{\keywords}[1]{\textbf{Keywords ---} #1}

\definecolor{Red}{RGB}{212,28,48}
\definecolor{Green}{RGB}{81,153,74}
\definecolor{Blue}{RGB}{0,128,255}
\definecolor{Yellow}{RGB}{255,204,0}

\begin{document}
\bibliographystyle{plainnat}

\title{GPU Fast Convolution via the Overlap-and-Save Method in Shared Memory}

\author[1]{Karel Ad\'{a}mek}
\author[2]{Sofia~Dimoudi}
\author[3]{Mike~Giles}
\author[1]{Wesley~Armour \thanks{E-mail address: \texttt{wes.armour@oerc.ox.ac.uk}} }
\affil[1]{Oxford e-Research Centre, Department of Engineering Sciences, University of Oxford, 7 Keble road, Oxford, OX1 3QG, United Kingdom}
\affil[2]{Centre for Advanced Instrumentation, Durham University, South Road, Durham, DH1 3LE, United Kingdom}
\affil[3]{Mathematical Institute, University of Oxford, Andrew Wiles Building, Radcliffe Observatory Quarter (550), Woodstock Road, Oxford, OX2 6GG, United Kingdom}

\maketitle

\begin{abstract}
We present an implementation of the overlap-and-save method, a method for the convolution of very long signals with short response functions, which is tailored to GPUs. We have implemented several FFT algorithms (using the CUDA programming language) which exploit GPU shared memory, allowing for GPU accelerated convolution. We compare our implementation with an implementation of the overlap-and-save algorithm utilizing the NVIDIA FFT library (cuFFT). We demonstrate that by using a shared memory based FFT we can achieved significant speed-ups for certain problem sizes and lower the memory requirements of the overlap-and-save method on GPUs.
\end{abstract}


\section{Introduction}
\label{sec:introduction}
Convolution is one of the most fundamental signal filtering techniques, widely used in signal processing, to aid discovery in many areas of natural sciences. It is a linear operation involving an input signal $s$ of length $N_s$ and a response function (or a filter) $h$ of length $M$. There are two principal approaches to linear filtering, where their usability depends on the length of the response function $h$. 

When the filter ($h$) is short it might be beneficial to calculate convolution in the time-domain using the formula for discrete convolution
\begin{equation}
\label{eqa:timeconvolution}
y[n]=h[k] \star s[n]=\sum_{k=0}^{M-1}s[n-k]h[k]\,, 
\end{equation}
where $y[n]$ are elements of the filtered signal, and brackets $[\,]$ denote quantities that are discrete (sampled). The complexity of time-domain convolution is $O\big(N_s^2\big)$.

If we have a longer filter it might be better to invoke the convolution theorem and calculate convolution in the frequency-domain using a Fourier transformation. The convolution theorem states that \cite{LYONS:UDSP}
\begin{equation}
\label{eqa:convolutiontheorem}
h[k]\star s[n]=\mathrm{FT}^{-1}\left(H[m] \cdot S[m]\right)\,, 
\end{equation}
where $H=\mathrm{FT}(h)$ and $S=\mathrm{FT}(s)$ are Fourier pairs of $h$ and $s$ and $\mathrm{FT}$ and $\mathrm{FT}^{-1}$ is discrete Fourier transformation and its inverse respectively. By using Fourier transformation in the convolution calculation we are performing circular convolution (as opposed to linear convolution (eq. \ref{eqa:timeconvolution})), which introduces an aliasing effect, where samples at the edges\footnote{This depends on the character of the filter used. Filters that use only future samples will be aliased with the end of the segment, filters that use past samples will be aliased with the beginning of the segment, while a time centred filter introduces aliasing at both ends. The number of aliased samples is equal to the unpadded length $M$ of the filter.} of the input signal are added together rendering them useless for convolution. Therefore we have to pad both the filter and the input signal with zeros (called zero padding), to the same size of at least $0 \leq m<N_s+M-1$.

The convolution theorem allows us to replace convolution in the time-domain by point-wise multiplication in the frequency-domain. This, however, would not be computationally feasible without the Fast Fourier Transformation (FFT) algorithm, which decreases the cost of the discrete Fourier transformation to $O\big(N_s\log_2\left(N_s\right)\big)$. Using the FFT algorithm and the convolution theorem to perform convolutions is often called \textit{fast convolution}. 

Determining when to use time-domain convolution as opposed to frequency-domain convolution depends on many factors including the character of the problem being solved, implementation, the hardware used etc.

As mentioned above, frequency-domain convolution requires that the input signal and the filter are both of the same length. To calculate the convolution of a long input signal in the frequency-domain, we have to perform long FFTs on both. This can be very inefficient in terms of computations and memory storage, particularly if we are applying multiple filters. Two commonly used algorithms to overcome these shortcomings are the \textit{overlap-and-save} (OLS) or \textit{overlap-and-add} (OLA) \cite{press1992num} methods. 

The overlap-and-save(add) is a hybrid method which combines advantages of time-domain convolution with frequency-domain convolution. It allows us to break the input signal into segments of length $N$ and use fast convolution independently on each segment. The two methods differ in the way they deal with aliased samples and how the output is constructed. The overlap-and-save method discards the aliased samples from each segment and saves only the correct part of the segment to an appropriate place in the output signal. The overlap-and-add method adds together aliased samples from the neighboring segments to create the correct output. Therefore a parallel implementation of the overlap-and-add method requires exclusive access to the areas of memory that contain the aliased output signal. 

The fast convolution, which is performed on each segment, has four steps: forward FFT of a segment; point-wise complex multiplication of the filter and the segment in frequency-domain; inverse FFT of the convolved segment; and rejection of the edges. These steps are traditionally performed using libraries or custom code, with the input and output stored in the GPU device memory\footnote{Device memory (sometimes called main memory or global memory) has the lowest memory bandwidth on the GPU and as such takes the most time to access} for each step. This is a limiting factor when considering the convolution of the segment as a whole.

The novelty of this work and its focus is to enable fast convolution by storing signal segments and filters in the fastest areas of GPU memory. Performing the convolution and the associated inverse FFT on data held in these fast memories allows us to eliminate device memory traffic and hence accelerate the convolution algorithm on GPUs. 

The novelty of this work and its focus is to enable fast convolution by exploiting the fastest areas of GPU memory, registers and shared memory. To do this we needed to write FFT codes that will operate directly on data stored in shared memory (NVIDIA library functions do not do this). Using these codes we are able to perform the convolution and the associated forward and inverse FFT on data held in the fastest areas of GPU memory and hence accelerate the convolution algorithm on GPUs. Specifically, we can eliminate expensive access to the device (global) memory which is otherwise required. With this goal in mind, we have implemented a basic version of the Cooley-Tukey FFT algorithm \cite{Coo-Tuk:1965:FFT} for complex-to-complex FFTs and a basic version of the Stockham FFT algorithm \cite{Stockham:1967:FFT} for real-to-complex and complex-to-real FFTs. We have implemented these FFT algorithms so that they can execute on data held in shared memory\footnote{Shared memory is a small but fast area of GPU memory and can be treated as a user managed cache}. The purpose of this work is to demonstrate the viability of our approach of moving operations into GPU kernels using device ready algorithms. The choice of the optimal FFT algorithm and implementation of optimized and efficient FFT algorithms on GPUs is beyond the scope of this work but will serve as a focus of our future work.

We have chosen to focus only on the overlap-and-save method rather than on the overlap-and-add method because the overlap-and-add method would require a synchronization step between segments due to a race condition that would occur when neighbouring segments try to write their computed data to the output signal stored in GPU device memory.

The work presented in this paper was developed for NVIDIA GPUs, therefore we have used the CUDA language extension for our work. The investigation of OpenCL or any other framework is outside the scope of this work. This work has been used to enable real-time processing of time-domain radio astronomy data \cite{Sofia:2018:FDAS, 2017arXiv171110855A, 2015arXiv151107343D}.

Our GPU implementation of the overlap-and-save method with a basic user interface is available on GitHub\footnote{\url{https://github.com/KAdamek/GPU_Overlap-and-save_convolution}}. The user interface we provide allows the user to test the functionality of our implementation. A more detailed description is provided on our Github wiki.

\section{Related work}
The comprehensive study of the convolution algorithms on CPUs, GPUs and FPGAs was conducted by \citet{Fowers:2013:ConvolutionsComparison}. They have compared convolution algorithms by their computational cost, energy efficiency and execution time for a range of input signal sizes and filter lengths. Their investigation shows that the time-domain convolution is faster for either short filters or short input signals. For longer input signals and longer filters, it is beneficial to use the overlap-and-save method. The performance of the  NVIDIA cuDNN library, in the context of convolutional neural networks, was investigated by \citet{Jorda:2019:cuDNNperf}. The authors present different algorithms used by the cuDNN to calculate two-dimensional convolution. Although this is for two-dimensional convolutions it shows the advantage of frequency-domain convolution for larger filters and input signals.

Both overlap-and-save (or OLA) and FFT algorithms are well known and extensively researched, having lots of coverage in literature. Both OLS and OLA methods have been implemented on GPUs \cite{Dobashi2013OaS, Lavin2015CNN}. The theory of these methods is also actively developed, for example \cite{Wefers2013Conf, Narasimha:2006:ModOLS, Fernandez:2013:MultimdimOLS} and references within.

The FFT algorithm and its implementation on GPUs is equally well researched and extensive publications can be found on the subject, for example \cite{Moreland:2003:FG:844174.844191, Govindaraju2008MicrosoftFFTs, Gutierrez:2008:FFT, fbfft:2014, volkov2008fitting, Yang:2014:oldFFT}. \citet{Govindaraju2008MicrosoftFFTs} focused on providing a set of FFT routines which would be applicable to a wide range of input signal lengths. The authors have used the Stockham algorithm to avoid reordering of the elements which is required when the Cooley-Tukey algorithm is used. \citet{Gutierrez:2008:FFT} deals with longer FFT from the host perspective with emphasis on long input signals. They have implemented the decimation-in-time Cooley-Tukey algorithm where part of the FFT is performed in shared memory and \citet{Moreland:2003:FG:844174.844191} described the implementation of the two-dimensional FFT real-to-real algorithm for image processing. More on FFTs in general can be found in \cite{Loan:1992:FFT_Computational_Frameworks}. 

There is also a number of GPU FFT source codes available  \cite{volkov2008fitting, fbfft:2014, Yang:2014:oldFFT}. However, these FFT codes were not suited for our needs for integration into the overlap-and-save method. The primary reason for this is that these FFT codes were not designed as device callable functions.

The FFT by Volkov, Kazian \cite{volkov2008fitting} stores larger FFTs (16 elements or more) using thread registers. Our implementation of convolution uses registers to store the values of the signal segment and current filter value. Further register utilization would lead to code slowdown.

The FFT code by Vasilache et al. \cite{fbfft:2014} focuses on FFT lengths that are too small for our intentions. The FFT length considered in the article is $N<256$. We require our implementation to work with the largest filters permitted by either shared memory\footnote{The size of shared memory ultimately limits the size of a signal segment that can be processed in our method.} or the number of active threads per thread-block. For example a filter size of 512 elements would require an FFT length of at least 1024 elements or longer.

Lastly the FFT code by Yang and Zhou \cite{Yang:2014:oldFFT} was written for the Fermi generation of GPUs and has not been updated for more modern GPU architectures.

Our FFT implementation differs from the previously published works because it is designed to use shared memory only and to be called from the GPU kernel itself. Therefore it deals only with short FFT lengths due to size limitation of the shared memory (currently N<=4096) and where N is a power of two. Moreover, our implementation of the Cooley-Tukey FFT algorithm cannot be used as a standalone FFT routine as it lacks element reordering which is not required for calculation of the convolution.


\section{Implementation}
We present our implementation of the overlap-and-save (OLS) method for NVIDIA GPUs using the CUDA programming language which uses a shared memory implementation of standard FFT algorithms to calculate one-dimensional convolutions. Our implementation of the OLS method can calculate complex-to-complex\footnote{Depending on the post-processing step this might be the complex-to-real convolution as well.} (C2C) and real-to-real (R2R) convolutions. These implementations are compared to an implementation of (direct) convolution that uses the NVIDIA cuDNN library and also to an implementation of the OLS method which uses the NVIDIA cuFFT library to perform the FFT parts of the OLS algorithm on the GPU.

In this section we describe all implementations used in this article starting with the NVIDIA cuDNN library \cite{cuDNNlib} implementation of convolution. Next, we describe the overlap-and-save method and its implementation using the NVIDIA cuFFT library \cite{cuFTTlib} (cuFFT OLS) which contains highly optimized and GPU ported FFT algorithms. Our implementation of the OLS method with shared memory FFT (SM-OLS) is described last. 

\subsection{Convolution via NVIDIA cuDNN}
\label{sec:cuDNN_OLS}
The NVIDIA CUDA Deep Neural Network library (cuDNN) is a GPU-accelerated library of deep neural networks primitives. The cuDNN library offers (among many other routines) forward convolution, which we have used as a comparison.

Our cuDNN convolution implementation is a real-to-real. The cuDNN library uses a range of different algorithms based on the task and the size of the input. We have left the cuDDN library to chose the most suitable convolution algorithm for our test case by using the flag \texttt{CUDNN\_CONVOLUTION\_FWD\_PREFER\_FASTEST}. Our tests are performed with one-dimensional data with a single channel\footnote{Channels in the context of cuDNN library are equivalent to the number of elements per structure in array-of-structures vs structure-of-arrays data layouts. Since we have a simple data layout, we have used the equivalent of structure-of-arrays.}, therefore we have used the \texttt{CUDNN\_TENSOR\_NCHW} data layout. Since we cannot be sure how many operations are performed by the cuDNN library we have not calculated the number of FLOPS for the cuDNN convolution implementation in our comparisons, instead we use the number of processed elements per second.

\subsection{Overlap-and-save Method}
We will first describe the common steps of the OLS method which are performed by all implementations. These steps apply to both C2C and R2R convolutions since both are performed in the Fourier domain which is complex.

\begin{figure}[htp]
\centering
\includegraphics[width=2.5in]{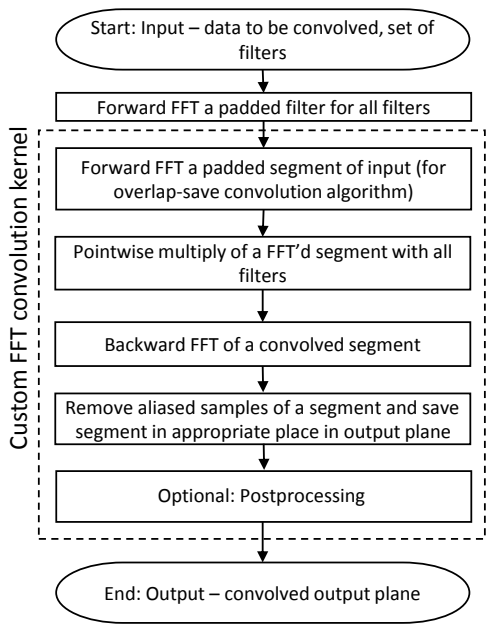}
\caption{Flow diagram of the overlap-and-save method: Our input is a signal which is to be convolved with a set of filters. The first step is to Fourier transform the padded filters. These will be used for convolution with each segment. In the next step, we separate the input signal into independent overlapping segments. The total overlap length for each segment is equal to the filter length, these segments need to be Fourier transformed. The third step is convolution in the form of complex point-wise multiplication. The convolved segment is then inverse Fourier transformed. In the last step we remove the aliased part of each segment and merge the clean parts to produce a continuous output. Optionally we can perform a post-processing step at the end.}
\label{fig:flowdiagram}
\end{figure}

\begin{figure}[!t]
\centering
\includegraphics[width=3.5in]{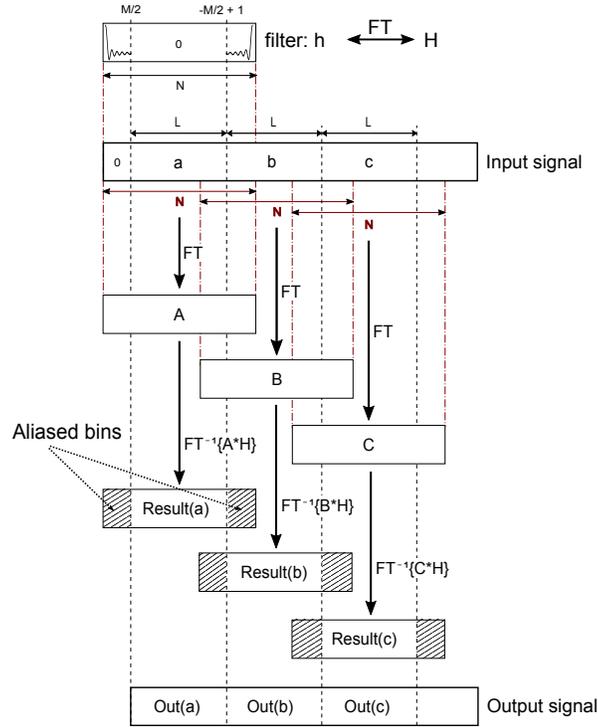}
\caption{Overlap-and-save method: The input signal, of length $N_s$, is separated into overlapping segments (A,B,C,...), where the amount of overlap is given by the filter length $M$. These segments are then processed independently, where $FT$ denotes Fourier transformation. At the end the aliased samples of the segment, which are equal to the filter length are discarded. This example uses a time centred filter, which aliases both ends of the segment.}
\label{fig:fig_overlap-save}
\end{figure}

A flow diagram of the overlap-and-save algorithm is shown in figure \ref{fig:flowdiagram} and the method is represented pictorially in figure~\ref{fig:fig_overlap-save}. 
We begin by separating the input signal of size $S$ into $N_\mathrm{seg}$ independent segments, all subsequent operations are then applied independently on each and every segment. Next a forward FFT is applied to each segment. What follows is the frequency domain convolution of the segment $A$ with every filter $f$ from $N_\mathrm{fil}$ filters, that is complex multiplication of the segment with one or more filters. After that, we apply an inverse FFT to the results and then discard the aliased edges of each block, recombining the samples from all blocks into the output. Optionally, we can apply some post-processing to the resulting output. In essence, this operation transforms a blocked circular convolution into one which is linear and continuous. 

In the overlap-and-save technique, (shown in figure~\ref{fig:fig_overlap-save}) the length of the segment, that is the FFT length, $N$ must be chosen such that it minimizes the fraction of discarded samples compared to the segment length. The number of discarded samples depends on the filter length $M$ that is being applied to the signal and are equal to $M-1$. Thus the number of correct (unaliased) samples in the segment is $L=N-M+1$. A higher fraction of discarded samples increases the overall number of segments required by the OLS method. To ensure good performance of the FFT algorithm on a signal segment we limit the segment length $N$  to be lengths equal to powers of two. The lengths of the segments from which we combine the convolved signal can be different for each implementation. The cuFFT-OLS performs better with longer segments while SM-OLS performs better with a shorter segment length. The convolved signal is not affected by the choice of the segment size.

\subsection{OLS method using cuFFT library}
\label{sec:cuFFT_OLS}
Using the cuFFT library, we have implemented one-dimensional convolution via the OLS method (cuFFT-OLS) for two variants of input data. We have implemented complex-to-complex and real-to-real convolutions. The pseudo-code for both variants of the cuFFT-OLS is shown in algorithm \ref{alg:cuFFT_OLS}. These variants only differ in the type of FFT used for the forward and the backward Fourier transform. Both using the cuFFT library to perform FFT routines.

\begin{algorithm}
 \SetAlgoLined
    \SetKwFunction{Boxcar}{Boxcar}
	\SetKwFunction{Decimate}{Decimate}
	\SetKwFunction{fFFT}{ForwardFFT}
	\SetKwFunction{bFFT}{InverseFFT}
	\SetKwFunction{removesamples}{RemoveAliasedSamples}
	\SetKwFunction{Sep}{Separate}
	\SetKwBlock{callback}{Callback begin}{end}
	\textbf{Input:} $x$, $f$\;
	\textbf{Output:} $y$\;
    \emph{Forward FFT of the filters}\;
    $F$ = \fFFT($f$)\;
    \emph{Separation of the signal into segments a, b, c, ...}\;
    ($a$, $b$, $c$, \ldots ) = \Sep($x$)\;
    \emph{Forward FFT of the individual segments}\;
    ($A$, $B$, $C$, \ldots ) = \fFFT($a$, $b$, $c$, \ldots )\;
    \callback{
        \emph{Per-element complex multiplication of the segment a with $N_\mathrm{fil}$ filters}\;
        \For{$s=0$ \KwTo $N_\mathrm{seg}$}{
            \For{$r=0$ \KwTo $N_\mathrm{fil}$}{
                $A[s]$ = $A[s]$ $\times$ $F[r][s]$\;
            }
        }
    }
    
    ($a$, $b$, $c$, \ldots ) = \bFFT($A$, $B$, $C$, \ldots )\;
    \callback{
        $y$ = \removesamples($a$, $b$, $c$, ...)\;
    }

	\caption{Pseudo-code for the cuFFT-OLS implementation. For the input we have input data $x$ and set of filters $f$. The output is the convolved result $y$. The FFT routines (\texttt{ForwardFFT} and \texttt{InverseFFT}) are either C2C for the complex input or R2C and C2R respectively for the real input.
	}
\label{alg:cuFFT_OLS}
\end{algorithm}

The most efficient way to implement cuFFT-OLS is to utilize a feature of the cuFFT library called callbacks. The cuFFT callbacks allow the user a per-element access to the data which are loaded or stored by the cuFFT routine and allow the user to perform pre- or post- processing of the data without any additional GPU kernels. 

We can use callbacks together with the forward FFT to perform frequency domain convolution (complex multiplication of the sample segment with the appropriate sample from multiple filters) and also with the inverse FFT where we can remove the aliased samples from the segment. While the latter eliminates problematic global memory access, the former callback has less effect. 

The callback used together with the inverse FFT means we do not need to store intermediate segments with aliased samples into the device memory. This is a significant bandwidth saving since the intermediate result is of size $N_\mathrm{seg}SF$ and it would have to be written to main memory (the output from cuFFT), then read so that aliased samples can be removed, then sorted as the final (corrected) output. 

The forward FFT callback eliminates proportionally only a small device memory access to the segments after the forward FFT. The main device memory access which stores the result of the frequency domain multiplication remains intact. Therefore the impact of this callback is marginal for a large number of filters.

The cuFFT library also allows the user to use some shared memory. The amount is however limited to 16kB which can accommodate only 2048 FFT elements while the optimal FFT length for cuFFT library is 8192 elements. Furthermore, this does not allow us to use forward and backward transform and as such does not remove problematic device memory access.

The disadvantages of the cuFFT-OLS implementation are that it has to load and store intermediate data to the device memory in between the frequency domain convolution (forward FFT step) and the inverse FFT. Another disadvantage is higher memory requirements as the last step (where we remove the aliased samples of the segments) cannot be performed in-place due to the non-deterministic nature of thread-block scheduling on GPUs. The advantage of the cuFFT implementation is that it works for any filter length and only relies on NVIDIA supported libraries.

\subsection{OLS method using shared memory FFT}
\label{sec:SM_OLS}
We present two versions of the one-dimensional overlap-and-save (OLS) method which is performed in the shared memory for NVIDIA GPUs using the CUDA programming language. The first implementation of OLS is for complex-to-complex\footnote{Depending on the post-processing step this might be complex-to-real convolution as well.} (C2C) convolutions, using a shared memory implementation of the Cooley-Tukey \cite{Coo-Tuk:1965:FFT} FFT algorithm. The second implementation of the OLS method is for real-to-real (R2R) convolutions. This implementation uses a shared memory implementation of the Stockham FFT algorithm \cite{Stockham:1967:FFT}. Our shared memory implementation of the OLS method follows the same steps as the cuFFT-OLS implementation, but has a significant difference, it incorporates all the steps required by the OLS method into one GPU kernel. This is possible because we can call forward and inverse FFT device functions directly from the GPU kernel, which eliminates the computationally costly device memory transactions, working instead on data held in shared memory and GPU registers. The pseudo-code for our shared memory OLS method is presented in algorithm \ref{alg:SMOLS}.

In our implementation of convolution through the OLS method in shared memory, each thread-block\footnote{A thread-block is a set of GPU threads which execute the same code and can cooperate using shared memory.} is assigned to one segment of the input data. Each thread-block applies a shared memory forward FFT and stores segment samples, which are now in the frequency domain, into registers. Each thread from the thread-block works with four samples. These segment samples are reused throughout the execution of the thread-block. Stored segment samples are then complex multiplied with appropriate samples from one or more filters. These filters are already in the frequency domain since they were Fourier transformed before thread-block execution. When the complex multiplication step is finished, the resulting samples are brought back to the time domain by applying an inverse FFT in shared memory and aliased samples are removed before storing them to the device memory. This ensures high data reuse of both segment and filter samples.


\begin{algorithm}
 \SetAlgoLined
    \SetKwFunction{Boxcar}{Boxcar}
	\SetKwFunction{Decimate}{Decimate}
	\SetKwFunction{fFFT}{ForwardFFT}
	\SetKwFunction{bFFT}{InverseFFT}
	\SetKwFunction{removesamples}{RemoveAliasedSamples}
	\SetKwFunction{thread}{threadId}
	\SetKwFunction{block}{blockId}
	\SetKwFunction{Sep}{Separate}
	\SetKwBlock{CUDABlock}{GPU kernel begin}{end}
	\textbf{Input:} $x$, $f$\;
	\textbf{Output:} $y$\;
	t = \thread\;
	b = \block\;
    \emph{Forward FFT of the filters}\;
    F = \fFFT(f)\;
    \emph{Each thread-block process one segment}\;
    \CUDABlock{
        \emph{Reading signal segment}\;
        $a[t]$ = $x$[$bN_\mathrm{Seg}+t$]\;
        \emph{Forward FFT of the individual segments}\;
        $A$ = \fFFT($a$)\;
        \emph{Per-element complex multiplication of the segment a with F filters}\;
        \For{$r=0$ \KwTo $N_\mathrm{fil}$}{
            $A$[t] = $A$[t] $\times$ F[r][t]\;
            $a$ = \bFFT($A$)\;
            $y$ = \removesamples($a$)\;
        }
    }
	\caption{Pseudo-code for the shared memory OLS implementation. For input we have input data $x$ and set of filters $f$. The output is the convolved result $y$. The shared memory FFT functions (\texttt{ForwardFFT} and \texttt{InverseFFT}) are either Cooley-Tukey C2C FFT for the complex input or Stockham FFT R2C and C2R respectively for the real input.}
\label{alg:SMOLS}
\end{algorithm}

We have chosen different FFT algorithms for C2C and R2R OLS implementations. The Cooley-Tukey FFT algorithm is more suited to complex-to-complex convolutions because we can use the fact that, for a point-wise frequency domain convolution, the order of the data elements in the convolved arrays does not matter as long as the order of the elements is the same for both the input signal segment and the filter, provided that the inverse FFT can work with the same order of elements. In normal circumstances, the Cooley-Tukey FFT algorithm requires a reordering to take place on the input or output data, but when used in convolution we can forgo this step and save some execution time.

The Stockham FFT algorithm is used in order to facilitate real-to-complex and complex-to-real Fourier transformation \cite{press1992num} these require that the elements of the input and output of the FFT algorithm are in the correct order. The Stockham FFT algorithm is an auto-sort algorithm which satisfies this condition. Our shared memory implementation of the Stockham FFT algorithm is 30\% slower on average than our shared memory implementation of the Cooley-Tukey FFT algorithm without the reordering step. This performance penalty is redeemed by the fact that for real-to-complex and complex-to-real Fourier transformations we can use an FFT length of half the size (compared to a C2C FFT) as described in \cite{press1992num}.

The benefit of having one GPU kernel is not only eliminating device memory accesses, but it also lowers memory requirements because we do not need to store intermediate results as with the cuFFT-OLS implementation. The disadvantage of this approach is that it works well only for small filter sizes $M \lesssim 3300$ (for Titan V GPU). This limitation is imposed by the size of the GPU shared memory.

The analysis of the SM-OLS GPU kernel reveals that it is limited by the shared memory bandwidth. For R2R version the kernel utilizes around 75\% of the shared memory bandwidth. The utilization is lower (50\%) for a segment size of 4096 elements. For the C2C version, the bandwidth utilization of the shared memory bandwidth is 50\%. This is in part because, for the first few iterations in the FFT routine, we use shuffle instructions which are not reflected by shared memory bandwidth utilization. The use of the shuffle instructions, however, increases utilization of the load-store instruction which is also high. The floating point (FP32) compute utilization is also high. The occupancy, a ratio of the maximum amount of active threads per streaming multiprocessor (SM) and active threads per SM, is only 50\%. This is a consequence of high register count used by the convolution kernel. The GPU registers are used to store the signal segment elements after forward Fourier transform which is reused and they are also used for storage of the currently processed signal segment which undergoes inverse Fourier transformation. The device memory bandwidth utilization ranges from 60\% down to 25\% for longer signal segments. The situation is similar for GPU kernels with non-local post-processing.


\section{Results}
	For our investigation we have used three NVIDIA GPU cards, the P100 GPU, the P4 GPU and the TitanV GPU (hardware specifications can be found in table \ref{tab:hardware}). 
	
\begin{table*}[htbp]
	\caption{GPU card specifications. The shared memory bandwidth is calculated as $\mathrm{BW (bytes/s)} = \mathrm{(bank\, bandwidth\, (bytes))} \times \mathrm{(clock\, frequency\, (Hz))} \times \mathrm{(32\, banks)} \times \mathrm{(\#\, multiprocessors)}$. We have used CUDA version 10.0.130 and cuDNN version 7.5.0. }
	\begin{center}
	\begin{tabular}{|l|r|r|r|}
	\hline
	\textbf{} & \textbf{P100} & \textbf{P4} & \textbf{TITAN V} \\
	\hline
	CUDA Cores & 3584 & 2560 & 5120 \\
	SMs & 56 & 20 & 80 \\
	Base/Max Core Clock & 1126/1303MHz & 810/1063MHz & 1220/1455MHz \\
	Memory Clock & 1406MHz & 6000MHz & 850MHz \\
	Gl. m. bandwidth & 720GB/s & 192GB/s & 652GB/s \\
	Shared m. bandwidth & 9121GB/s & 2657GB/s & 14550GB/s \\
	Memory size & 16GB &  8GB & 12GB\\
	TDP & 250W & 75W & 250W\\
	Max. sh. memory per thread-block & 48kB & 48kB & 48/96kB\\
	\hline
	\end{tabular}
	\label{tab:hardware} 
	\end{center}
	\end{table*}

	We have compared both shared memory implementations (C2C, R2R) of OLS convolution (SM-OLS) for several different filter and signal lengths and also for a varying number of filters with convolution implementations based on the cuDNN library and our implementation of the OLS method which uses cuFFT (cuFFT-OLS). For our results presented here, we have chosen to limit the input signal length to 2 million points or the number of filters to 8 (unless otherwise stated), in order to include the P4 GPU in our comparisons. The reason for this is that the P4 GPU card has a smaller device memory capacity and as such cannot process the same problem sizes as the P100 GPU or TitanV GPU.
    
    The length of the input signal or filter length, as well as the number of filters in our implementation, can be arbitrary and they are not limited to presented values. We have chosen the value of these quantities to present the scaling behaviour of the problem. The input signal length is arbitrary from the nature of the OLS method and limited only by available memory. The filter length is arbitrary, but in the case of the shared memory OLS it is limited by the maximum size of the FFT which can be processed (currently N=4096 points). The number of filters is also arbitrary and limited only by available memory.
    
\begin{figure}[htp]
	    \centering
	    \includegraphics[width=0.50\linewidth]{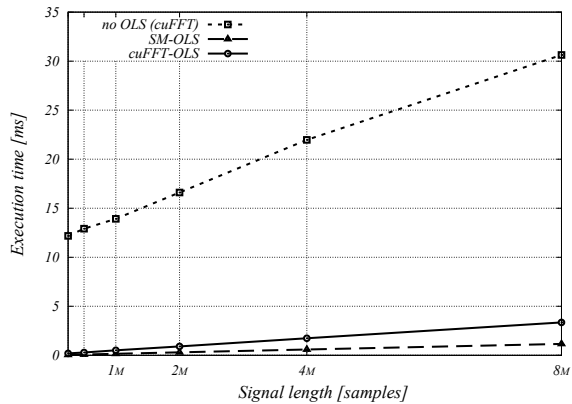}
	    \caption{Comparison of the execution time of convolution without OLS method using cuFFT, convolution via OLS method using cuFFT and convolution via custom FFT in shared memory. Results are for 8 filters of length 64 on TITAN V.}
    	\label{fig:fullconv}
    \end{figure}

    First, we have compared convolution without the OLS method using cuFFT, although OLS is a well established method this comparison shows how ineffective the standard convolution through the frequency domain can be for the case of convolution with multiple small filters. The execution time for convolution without OLS is presented in figure \ref{fig:fullconv}.
    
    \subsection{Comparison with cuDNN library convolution}
\begin{figure}[htp]
	    \begin{minipage}[t]{.485\textwidth}
            \centering
            \includegraphics[width=\linewidth]{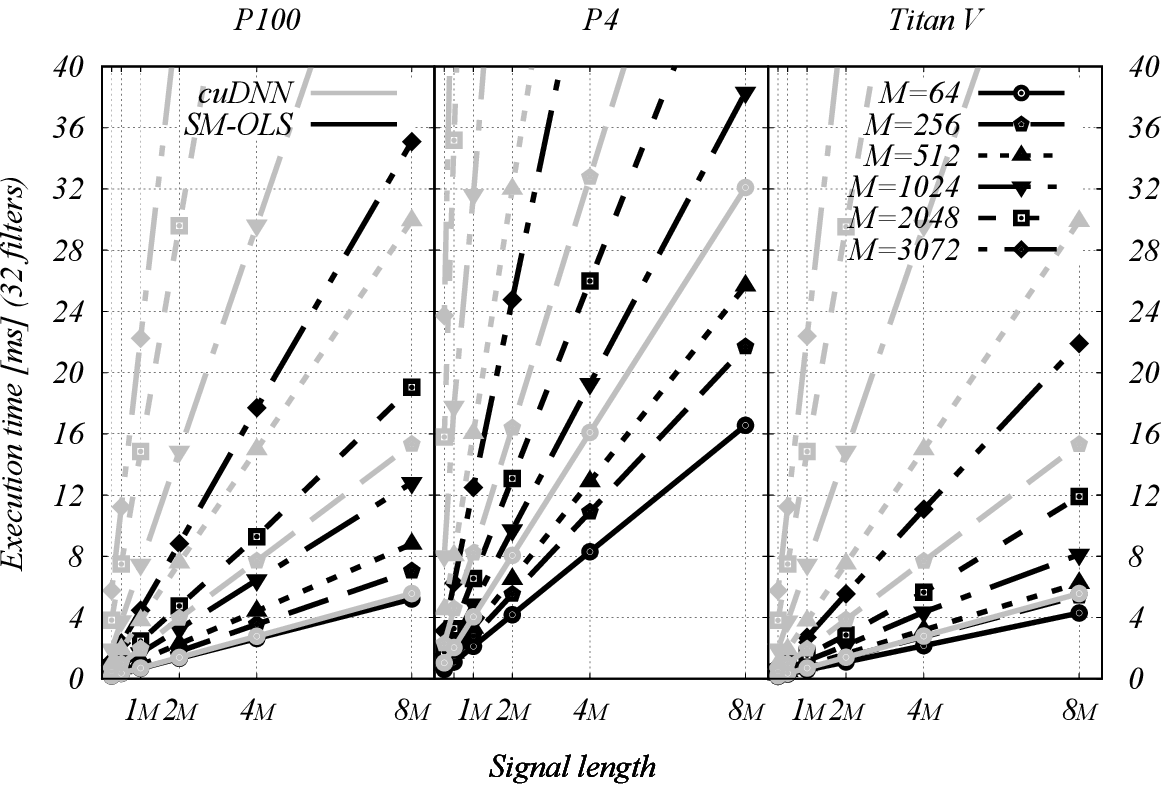}
	    \end{minipage}%
	    \hfill%
	    \begin{minipage}[t]{.485\textwidth}
		    \centering
    		\includegraphics[width=\linewidth]{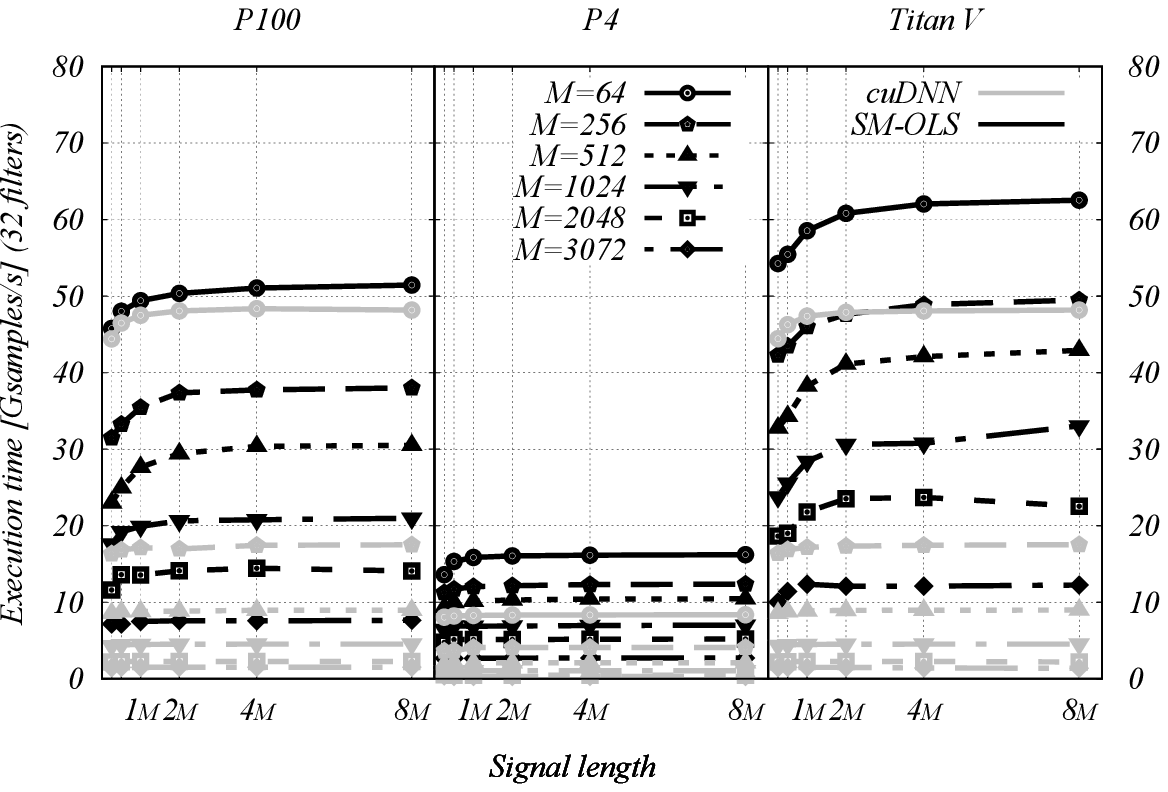}
	    \end{minipage}\\[0.2cm]
	    
	    \begin{minipage}[t]{.485\textwidth}
            \centering
            \includegraphics[width=\linewidth]{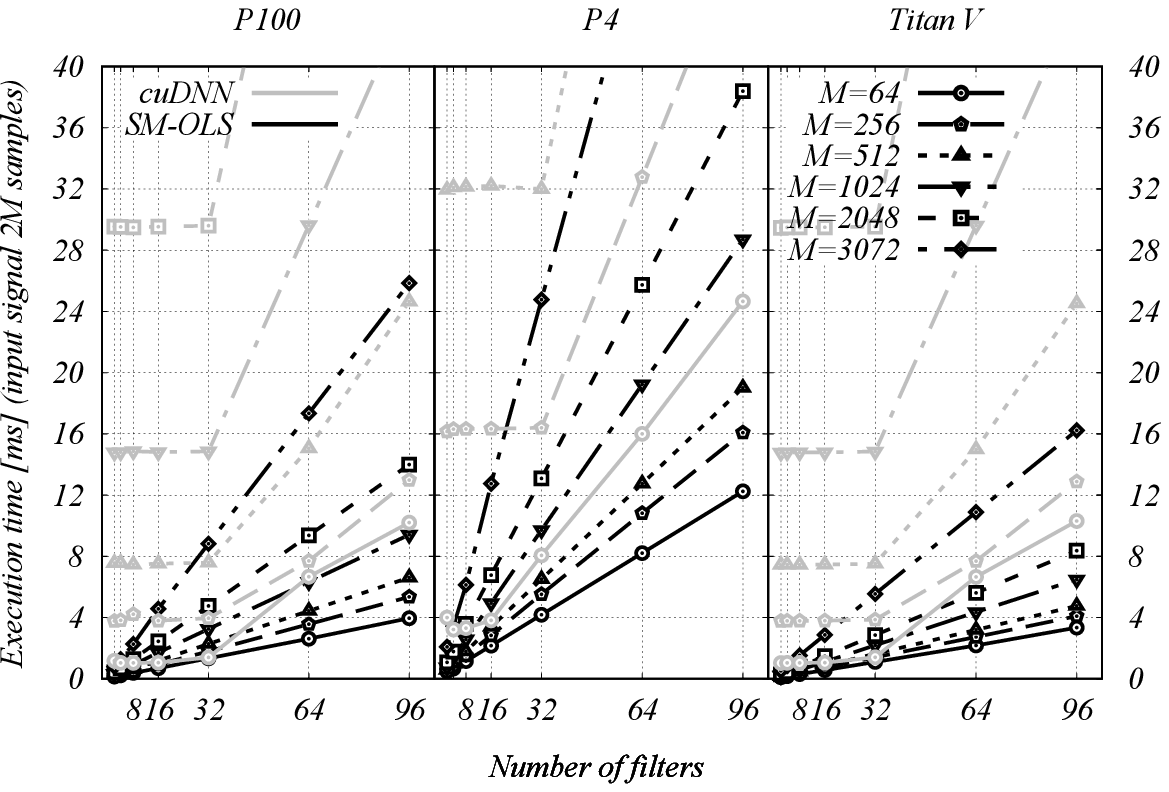}
	    \end{minipage}%
	    \hfill%
	    \begin{minipage}[t]{.485\textwidth}
		    \centering
		    \includegraphics[width=\linewidth]{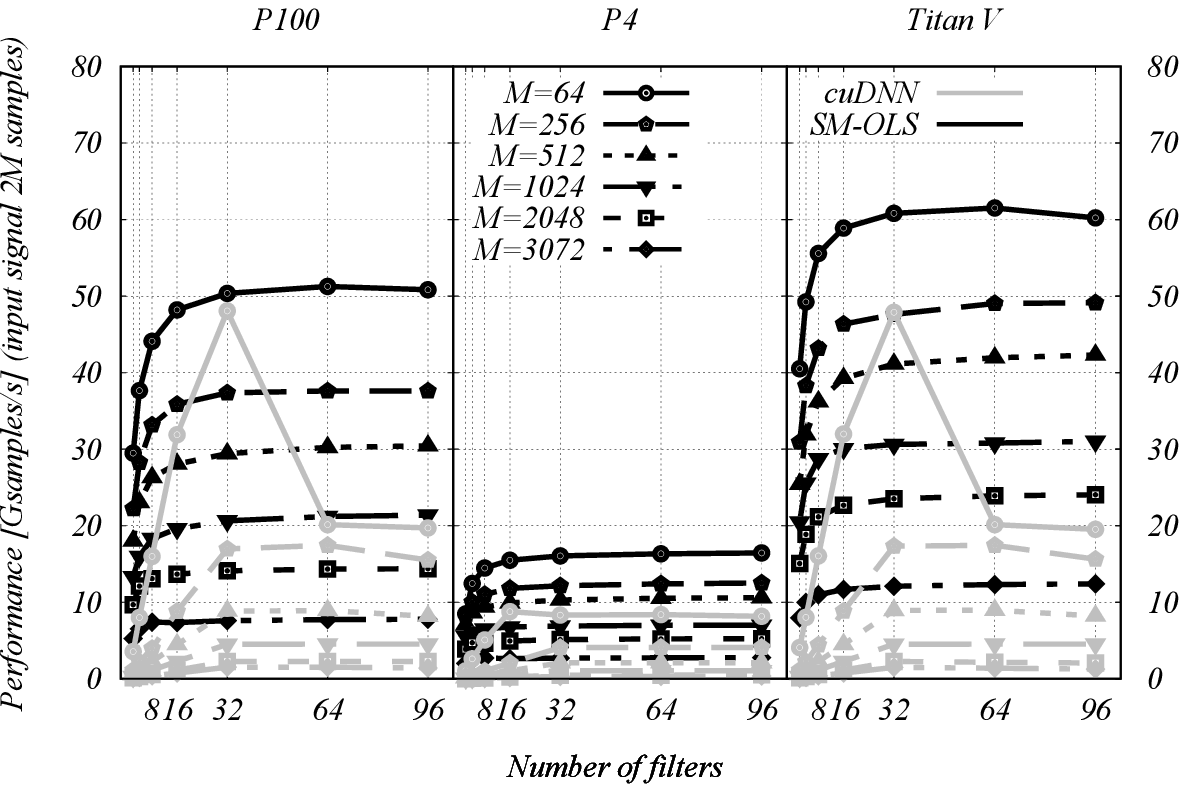}
	    \end{minipage}\\
	    \caption{The execution time of the R2R convolution on the left and the number of elements processed per second on the right via cuDNN (gray) and shared memory OLS (black) for different input signal lengths (top) and different number of filters (bottom).}
	    \label{fig:cuDNN_OLS_time}
    \end{figure}

\begin{figure}[htp]
	    \begin{minipage}[t]{.485\textwidth}
            \centering
              \includegraphics[width=\linewidth]{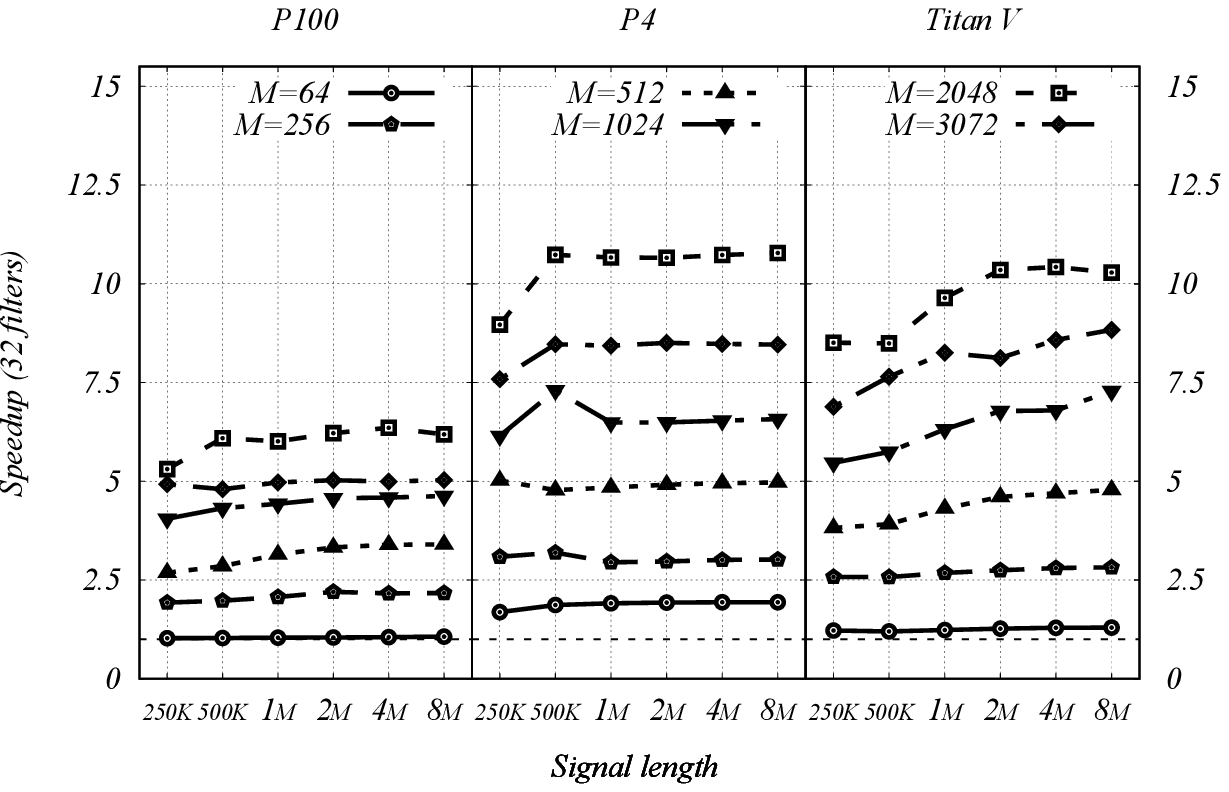}
	    \end{minipage}%
	    \hfill%
	    \begin{minipage}[t]{.485\textwidth}
		    \centering
		    \includegraphics[width=\linewidth]{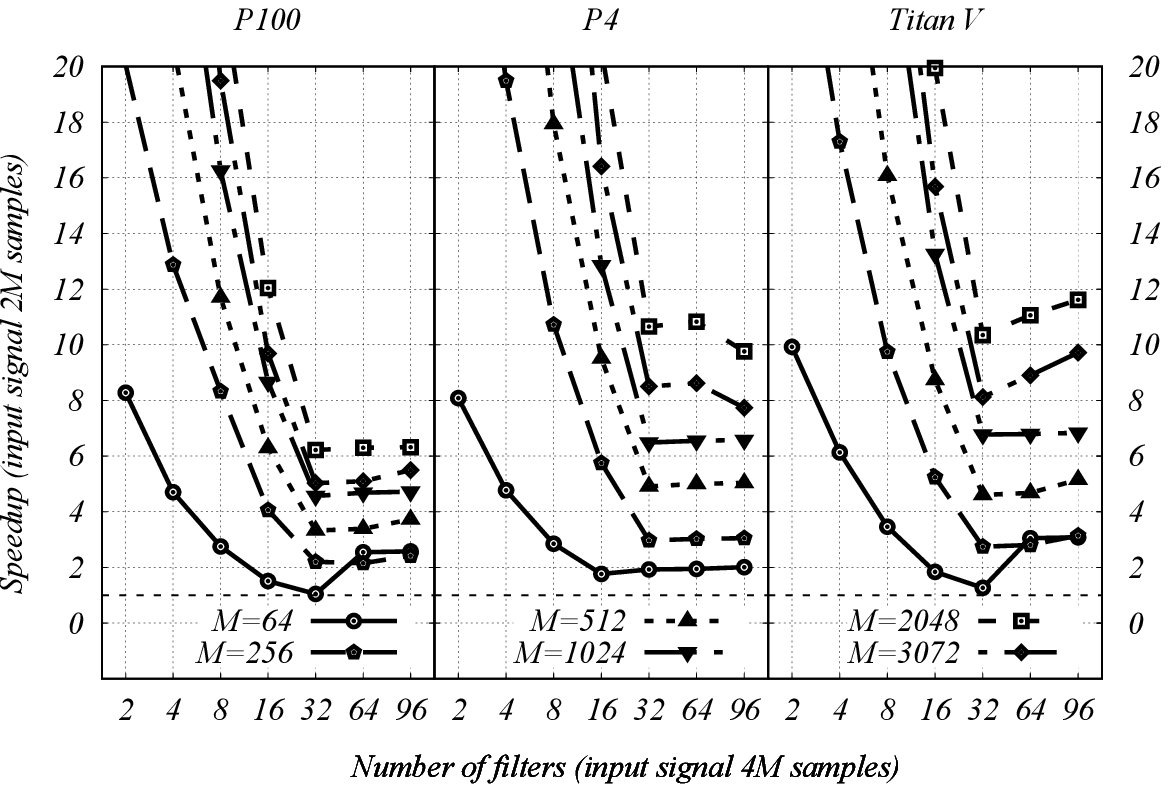}
	    \end{minipage}\\
	    \caption{The speedup factors of the R2R SM-OLS convolution with respect to the cuDNN convolution for different input signal lengths (left) and different number of filters (right).}
	    \label{fig:cuDNN_OLS_speedup}
    \end{figure}

    We begin by comparing the one-dimensional real-to-real SM-OLS convolution with one-dimensional real-to-real convolution via the cuDNN library. The execution time for the different input signal lengths and for the different number of filters is shown in Figure \ref{fig:cuDNN_OLS_time}. The speedup factor for the same configurations is shown in Figure \ref{fig:cuDNN_OLS_speedup}.
    
    \subsection{Comparison with cuFFT OLS convolution}
\begin{figure}[htp]
	    \begin{minipage}[t]{.485\textwidth}
            \centering
            \includegraphics[width=\linewidth]{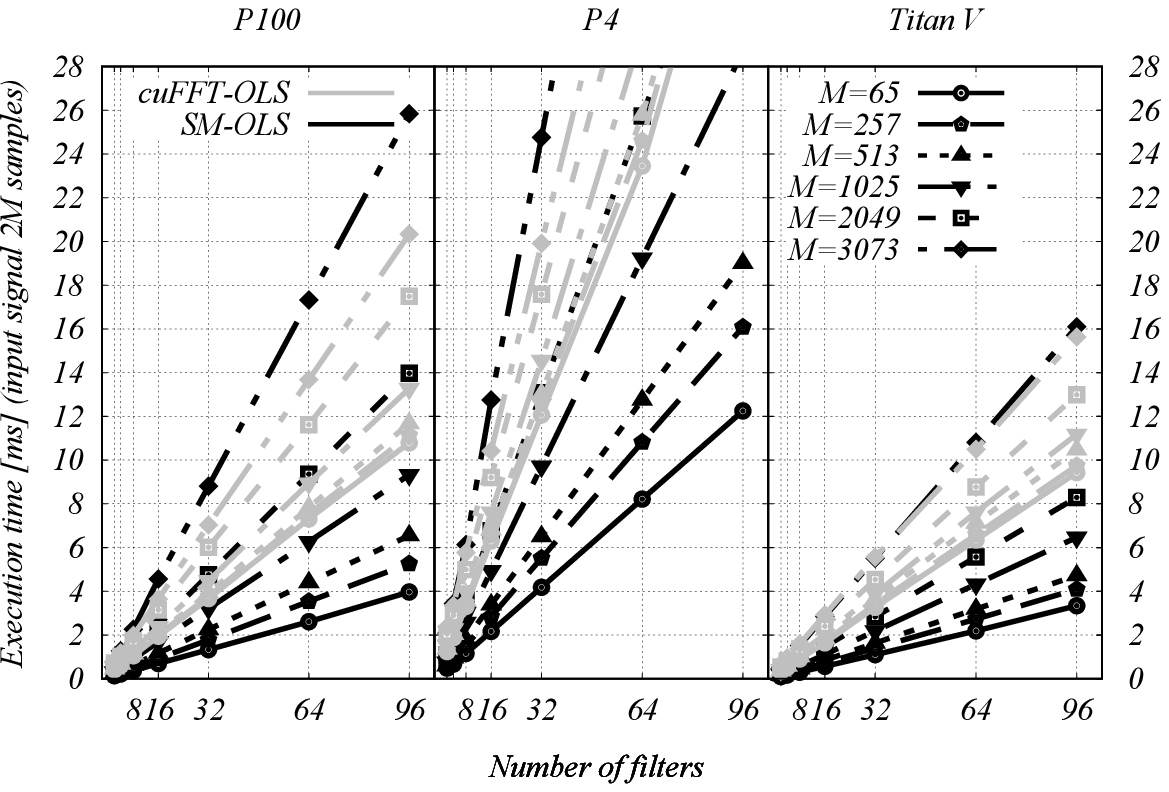}
	    \end{minipage}%
	    \hfill%
	    \begin{minipage}[t]{.485\textwidth}
		    \centering
    		\includegraphics[width=\linewidth]{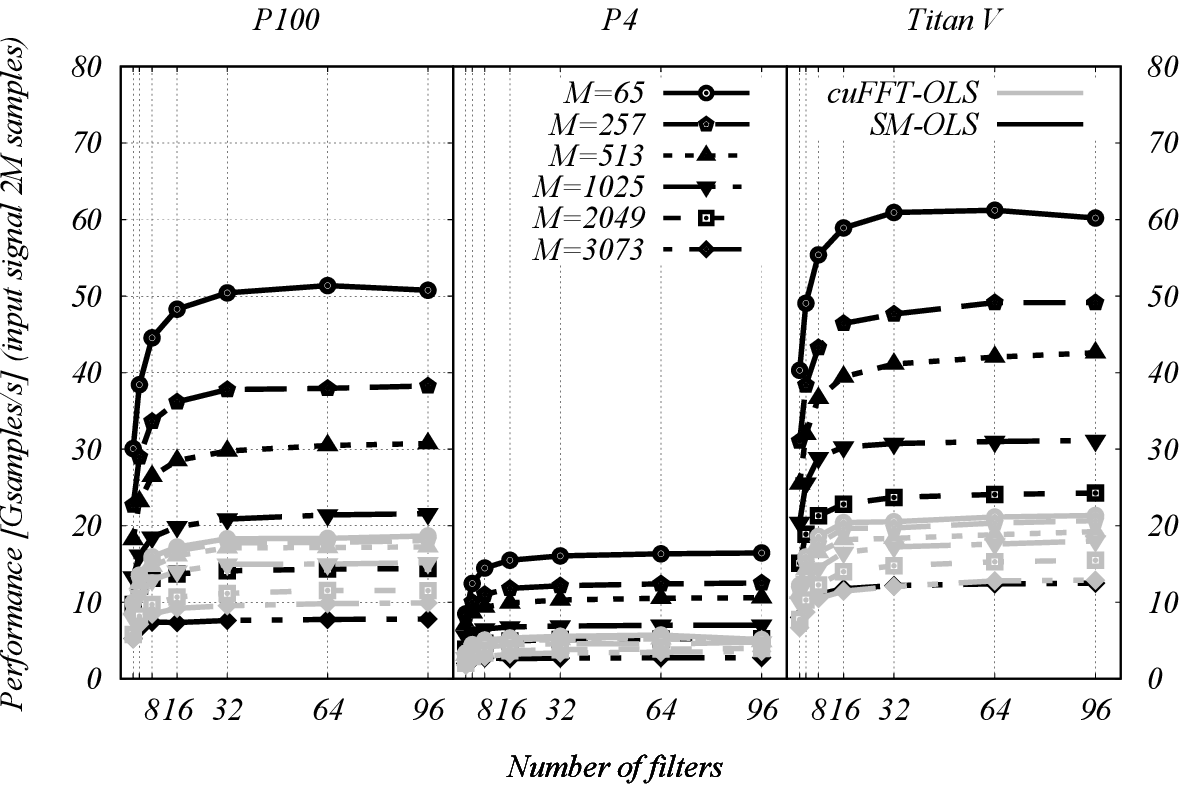}
	    \end{minipage}\\[0.2cm]
	    
	    \begin{minipage}[t]{.485\textwidth}
            \centering
            \includegraphics[width=\linewidth]{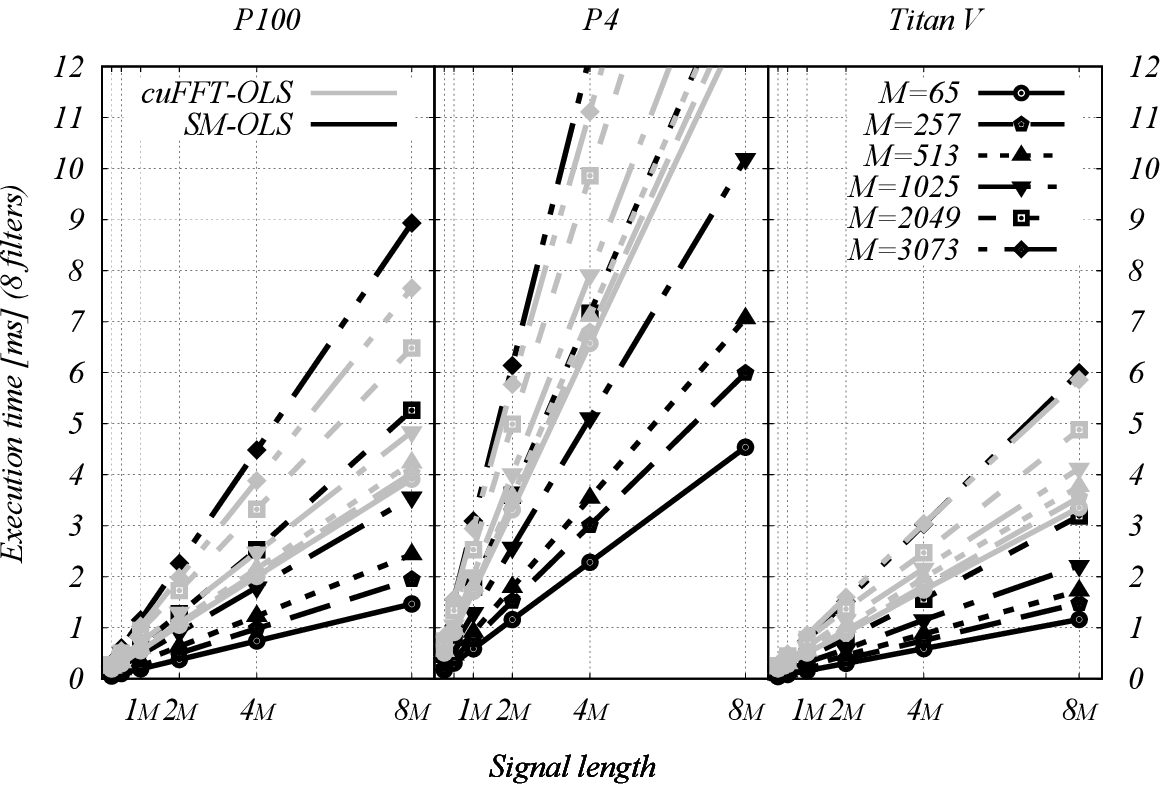}
	    \end{minipage}%
	    \hfill%
	    \begin{minipage}[t]{.485\textwidth}
		    \centering
		    \includegraphics[width=\linewidth]{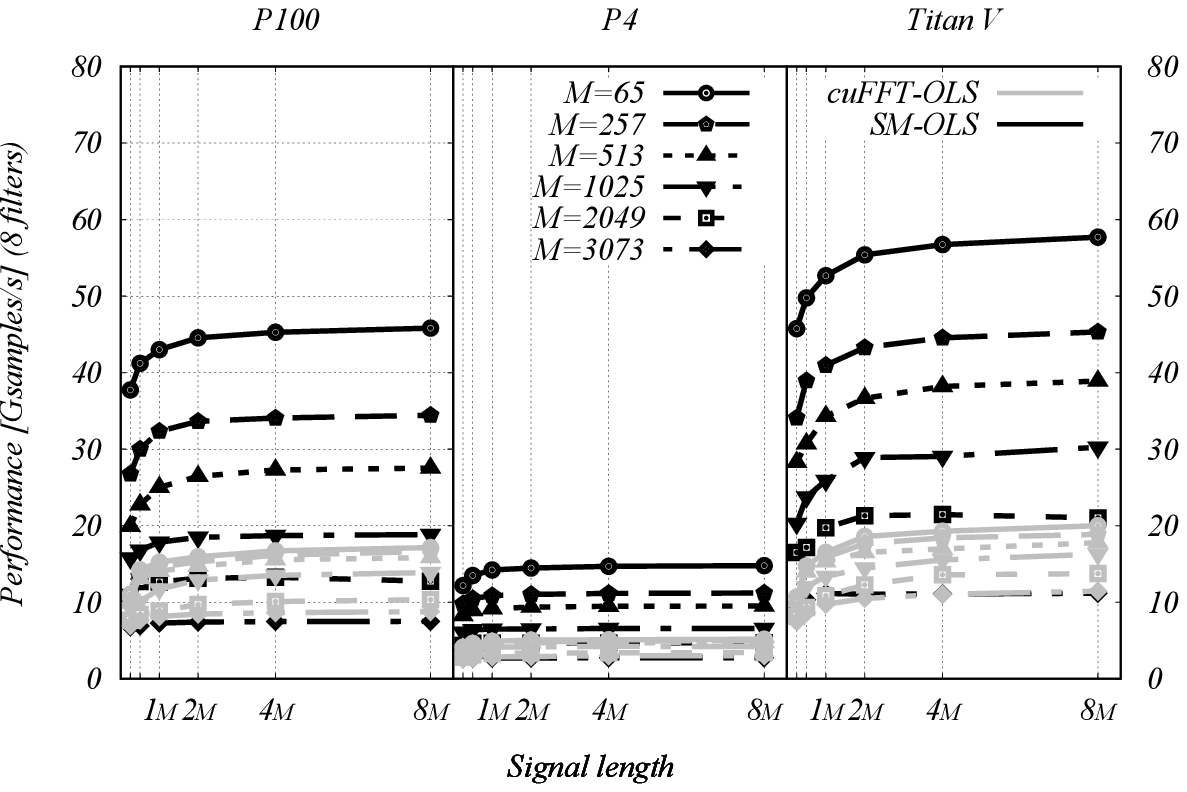}
	    \end{minipage}\\
	    \caption{The execution time of the C2C convolution on the left and the number of elements processed per second on the right of the SM-OLS convolution (black) and the cuFFT-OLS convolution (gray) for different number of filters (top) and increasing input signal length (bottom).}
	    \label{fig:C2C_time_flops}
    \end{figure}

    Next we present results for the comparison of complex-to-complex (C2C) Fourier domain convolution implementations. The execution time and the number of processed elements vs the number of filters, and vs input signal length is presented in Figure \ref{fig:C2C_time_flops}.  
    
\begin{figure}[htp]
	    \begin{minipage}[t]{.485\textwidth}
            \centering
            \includegraphics[width=\linewidth]{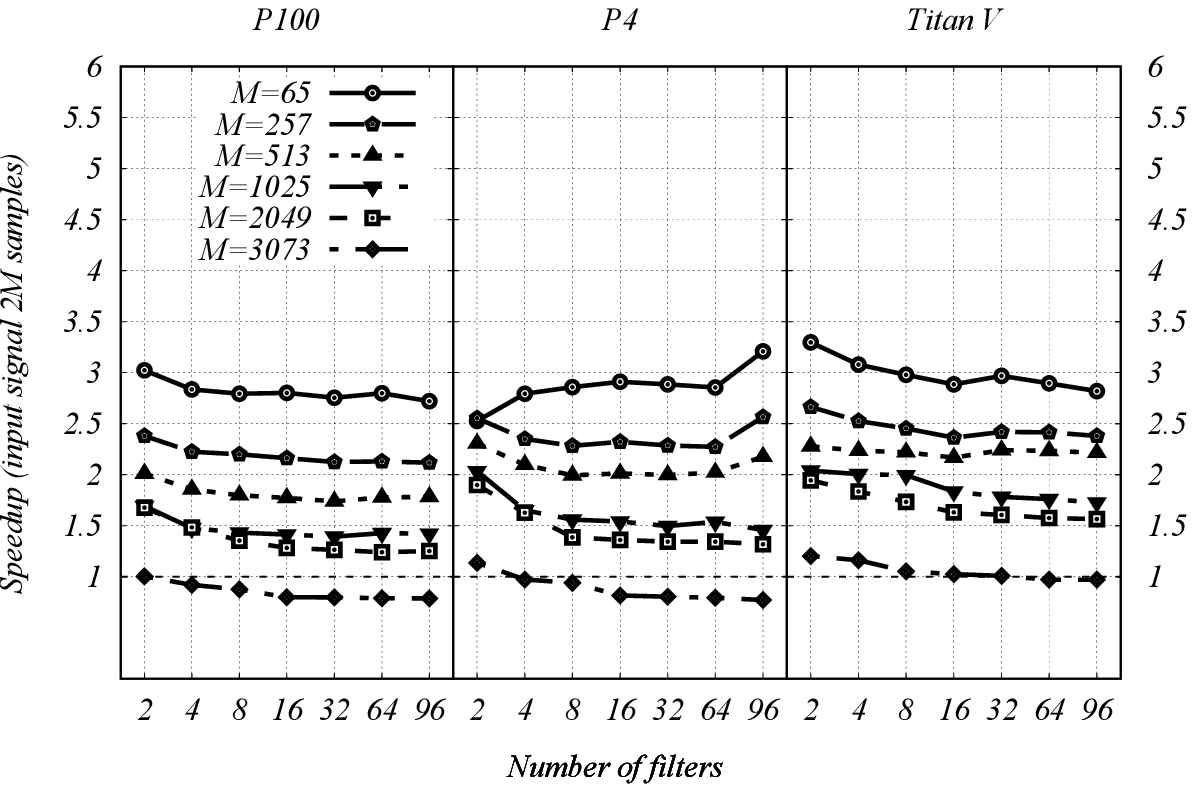}
	    \end{minipage}%
	    \hfill%
	    \begin{minipage}[t]{.485\textwidth}
		    \centering
		    \includegraphics[width=\linewidth]{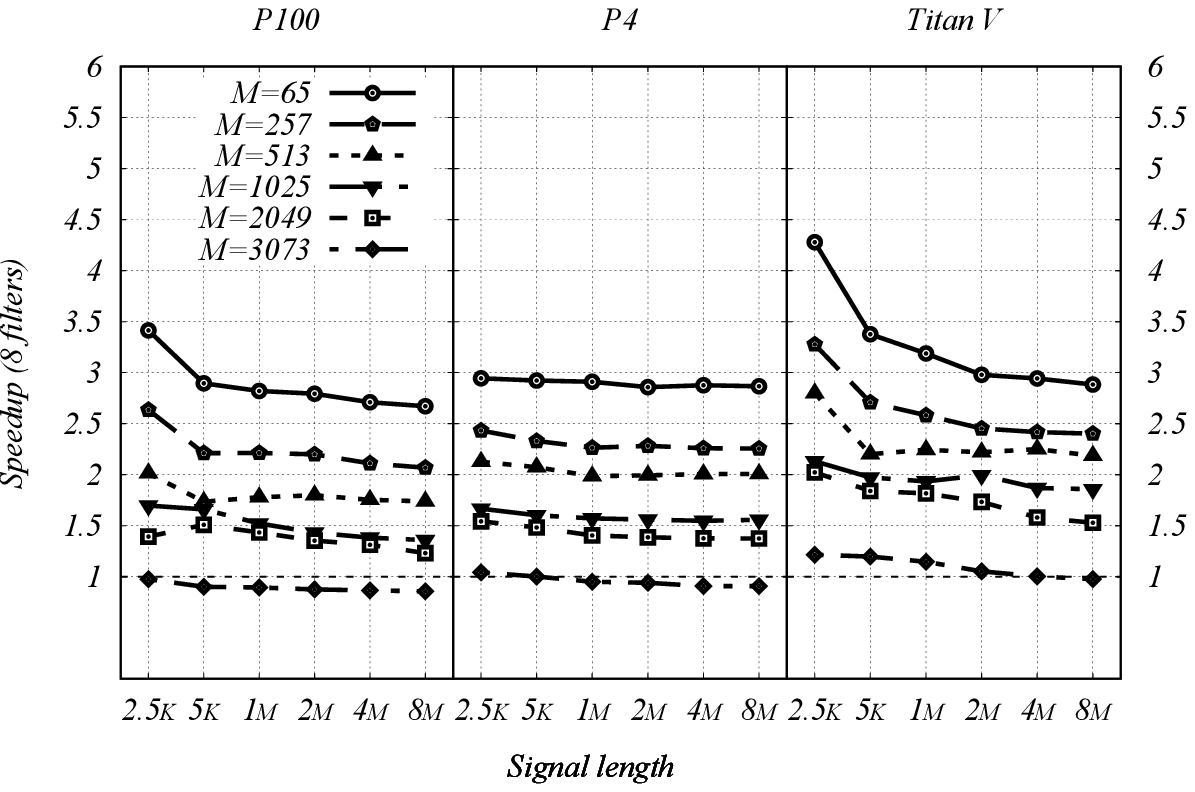}
	    \end{minipage}\\
	    \caption{The speed-up of the C2C SM-OLS convolution with respect to the C2C cuFFT-OLS convolution implementation for different filter lengths vs the number of filters (left), and vs the signal length (right).}
    	\label{fig:C2C_speedups}
    \end{figure}

\begin{figure}[htp]
	    \begin{minipage}[t]{.485\textwidth}
            \centering
            \includegraphics[width=\linewidth]{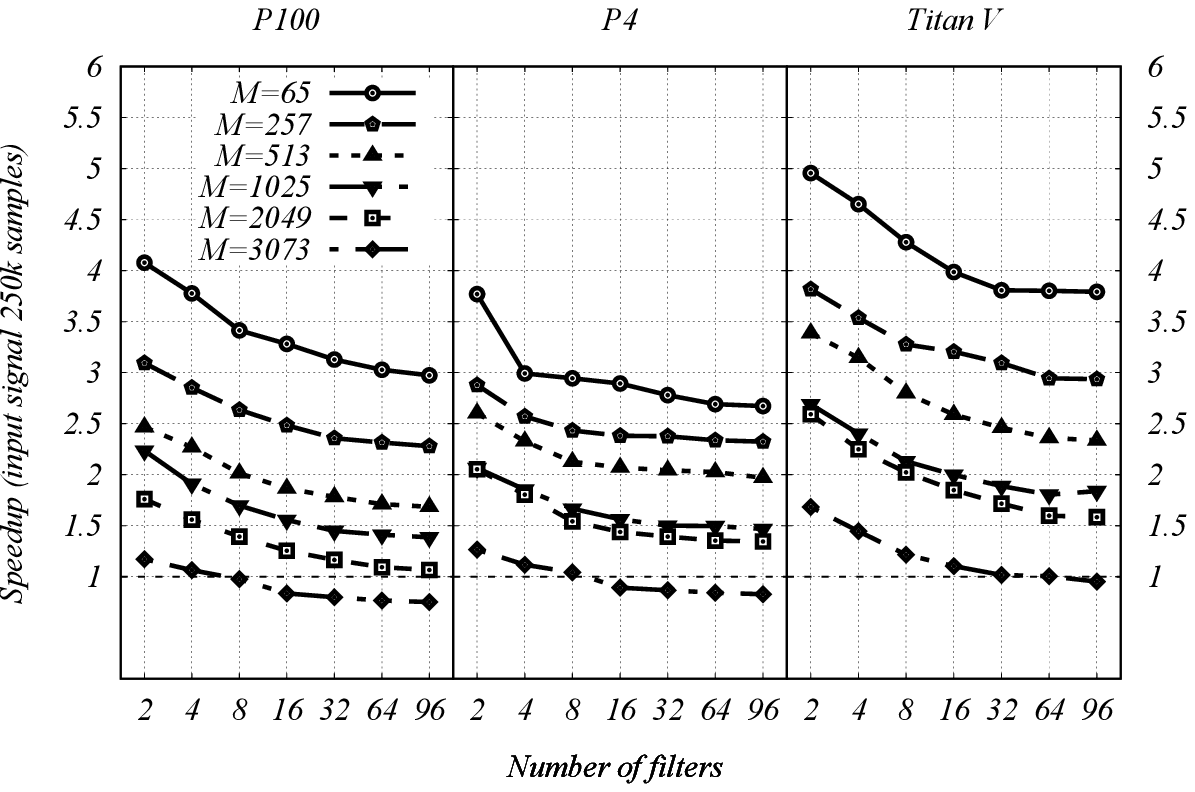}
	    \end{minipage}%
	    \hfill%
	    \begin{minipage}[t]{.485\textwidth}
		    \centering
		    \includegraphics[width=\linewidth]{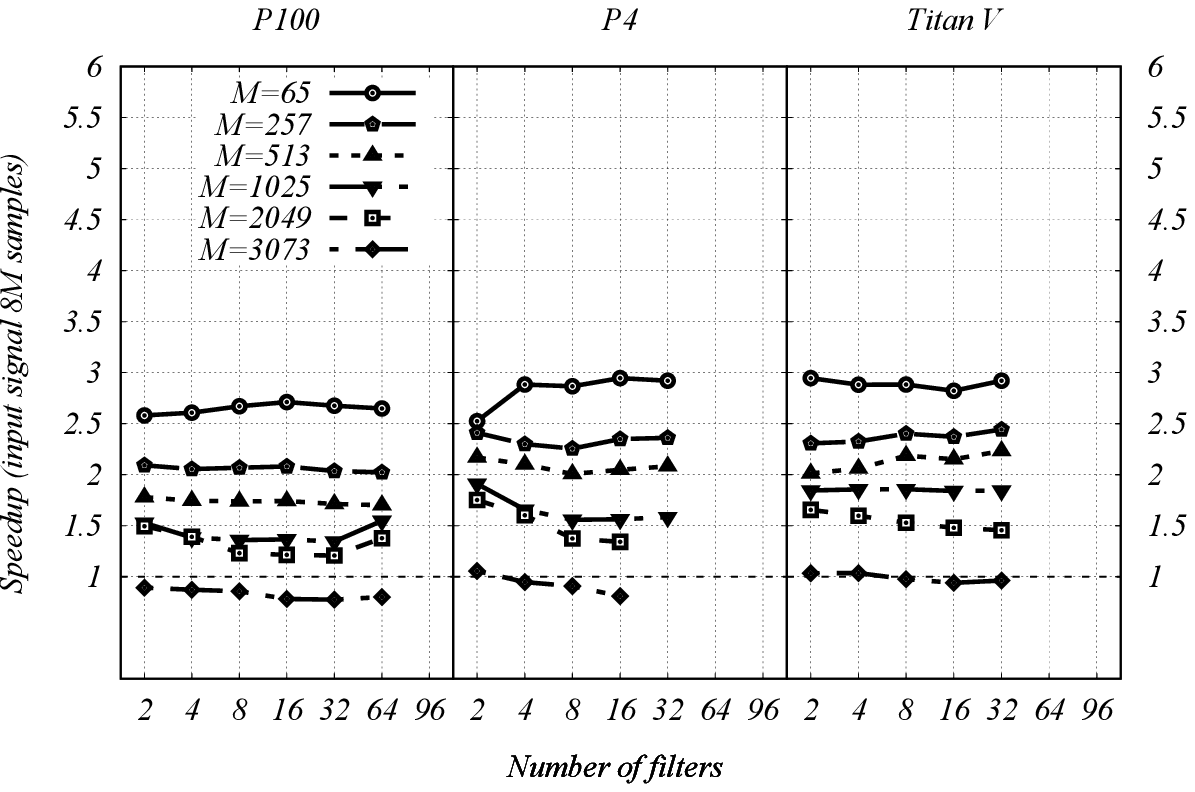}
	    \end{minipage}\\
	    \caption{The speed-up of the C2C SM-OLS convolution with respect to the C2C cuFFT-OLS convolution implementation for different filter lengths vs the number of filters (left) for signal lengths 250k and 8M samples. The number of filters is limited by amount of device memory the GPU has, this is why there are missing points for P4 GPU (8GB) and TitanV GPU (12GB).}
    	\label{fig:C2C_speedups_other_signallengths}
    \end{figure}

	The speed-up factors for different filter lengths vs the number of filters and vs the input signal length used are presented in Figure \ref{fig:C2C_speedups}. Furthermore, speedups for signal length other then 2M samples are shown in Figure \ref{fig:C2C_speedups_other_signallengths}.

\begin{figure}[htp]
	    \centering
	    \includegraphics[width=0.65\linewidth]{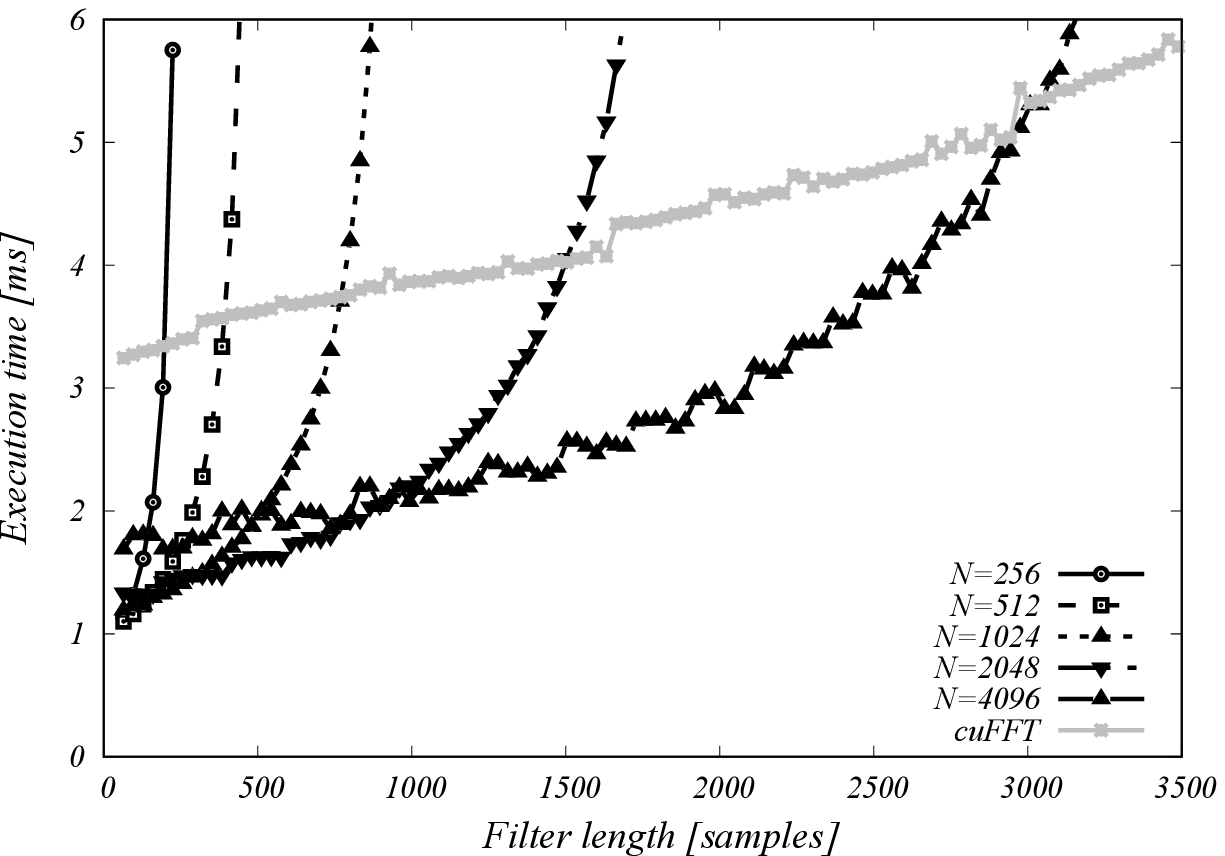}
	    \caption{The execution time of the SM-OLS convolution vs filter length for different segment (FFT) sizes. The execution time of the cuFFT-OLS convolution is added for comparison.}
	    \label{fig:C2C_time_conv_length}
    \end{figure}

	The cuFFT-OLS convolution performs best with segment size $N=8192$ for most of the filter sizes that we have investigated. The best performing segment size in the case of the SM-OLS convolution varies, this is because our FFT implementation performs better for smaller FFT lengths. Figure \ref{fig:C2C_time_conv_length} shows how the performance of the SM-OLS convolution depends on the  chosen FFT length (for TitanV GPU). Smaller FFT sizes become less effective with longer filter lengths, because the aliased part of the segment becomes a higher fraction of the overall FFT size and more segments are necessary to calculate the OLS convolution.

\begin{figure}[htp]
	    \begin{minipage}[t]{.485\textwidth}
            \centering
            \includegraphics[width=\linewidth]{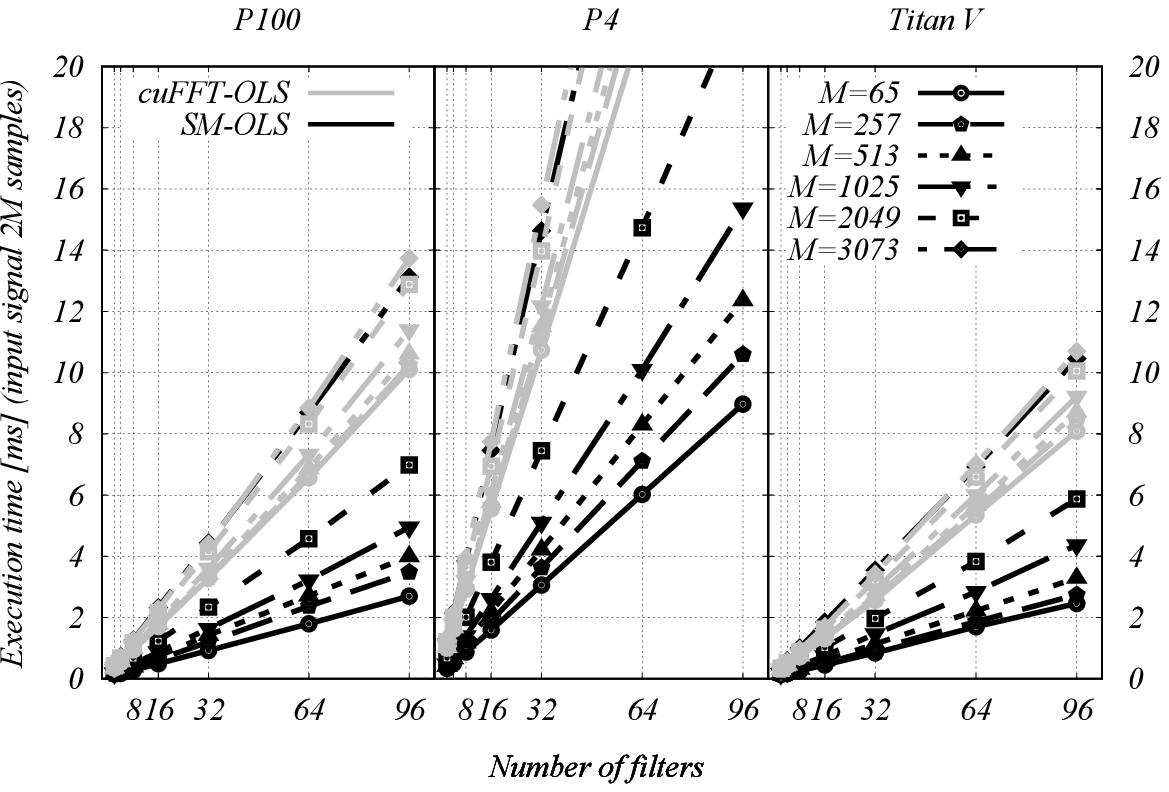}
	    \end{minipage}%
	    \hfill%
	    \begin{minipage}[t]{.485\textwidth}
		    \centering
    		\includegraphics[width=\linewidth]{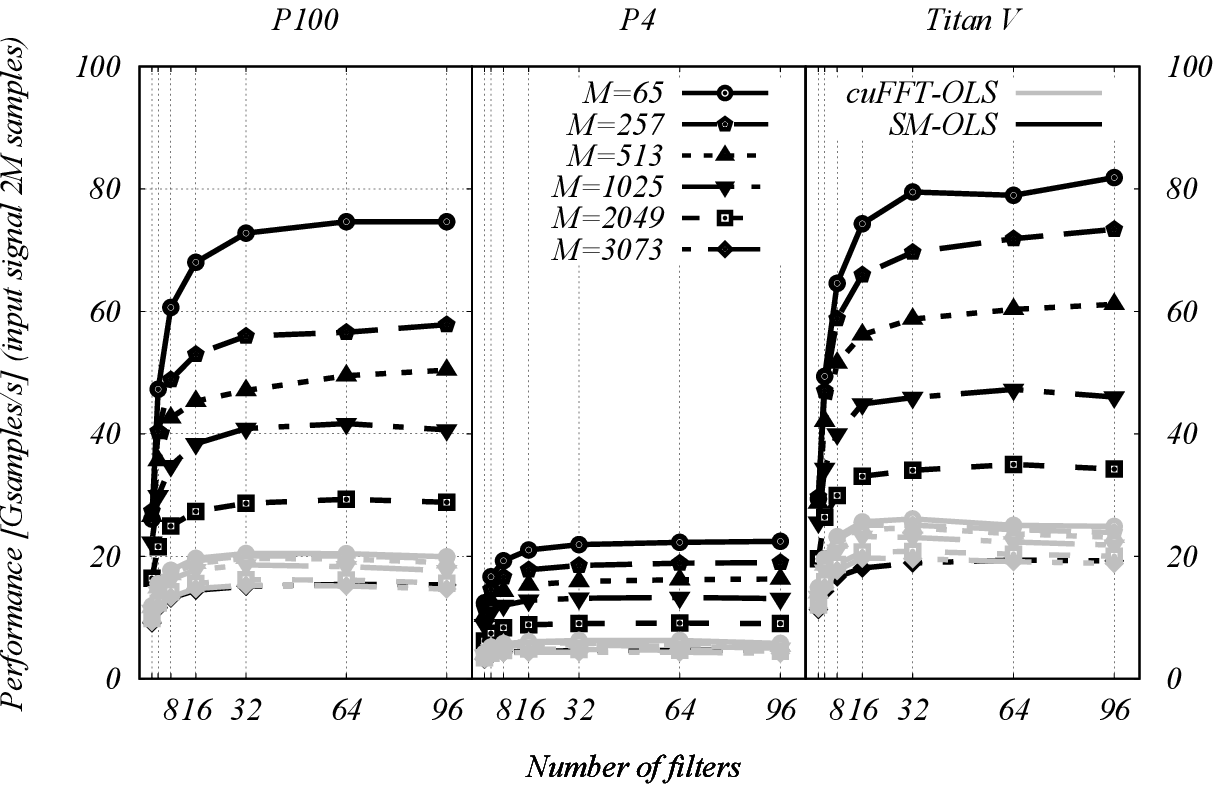}
	    \end{minipage}\\[0.2cm]
	    
	    \begin{minipage}[t]{.485\textwidth}
            \centering
            \includegraphics[width=\linewidth]{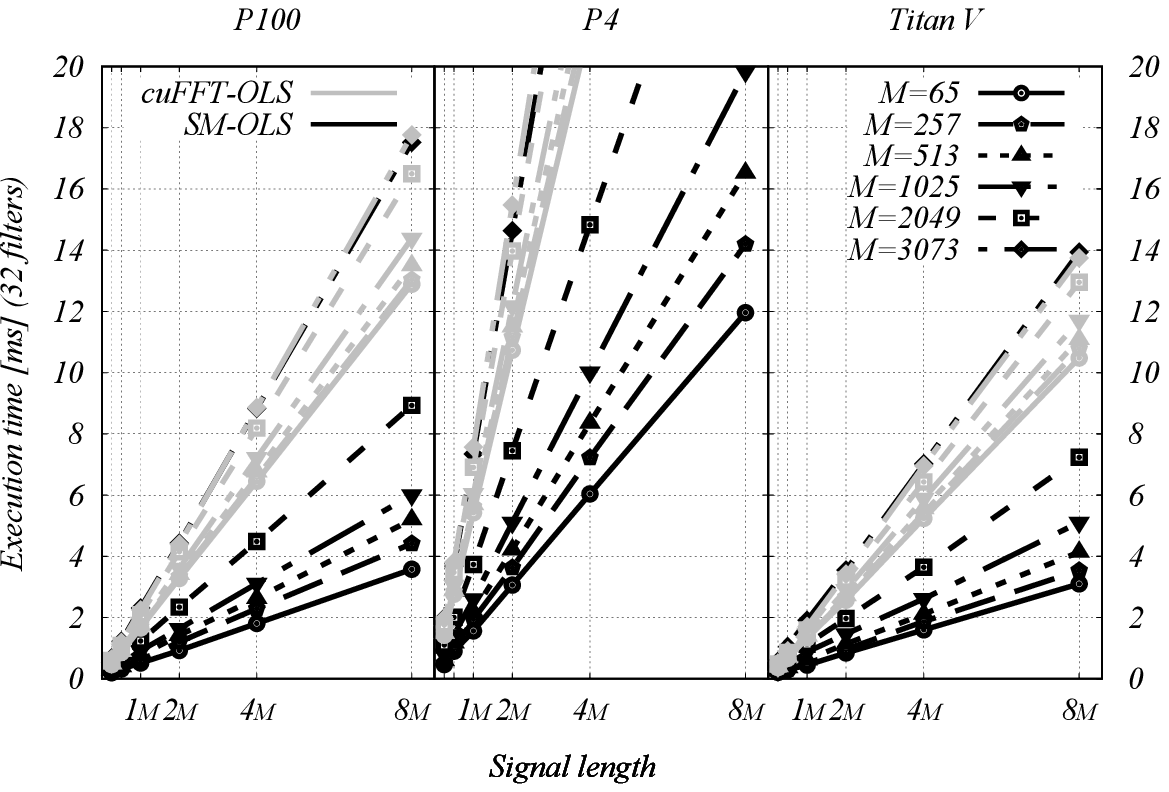}
	    \end{minipage}%
	    \hfill%
	    \begin{minipage}[t]{.485\textwidth}
		    \centering
		    \includegraphics[width=\linewidth]{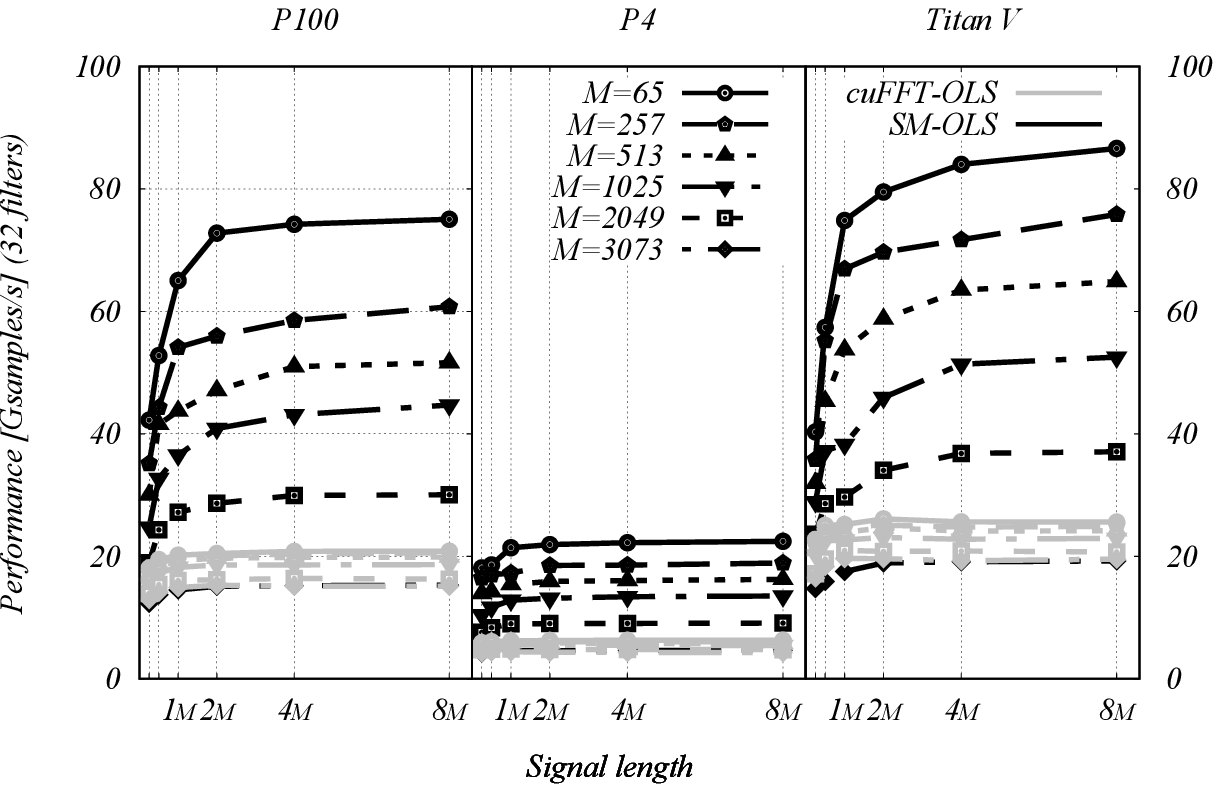}
	    \end{minipage}\\
	    \caption{The execution time of the R2R convolution on the left and the number of elements processed per second on the right of the SM-OLS convolution (black) and the cuFFT-OLS convolution (gray) for different number of filters (top) and increasing input signal length (bottom).}
	    \label{fig:R2R_time_flops}
    \end{figure}

	The comparison of real-to-real SM-OLS with cuFFT-OLS is similar. The execution time and the number of elements processed per second vs the number of filters, and vs input signal length is shown in Figure \ref{fig:R2R_time_flops}.
	
\begin{figure}[htp]
	    \begin{minipage}[t]{.485\textwidth}
            \centering
            \includegraphics[width=\linewidth]{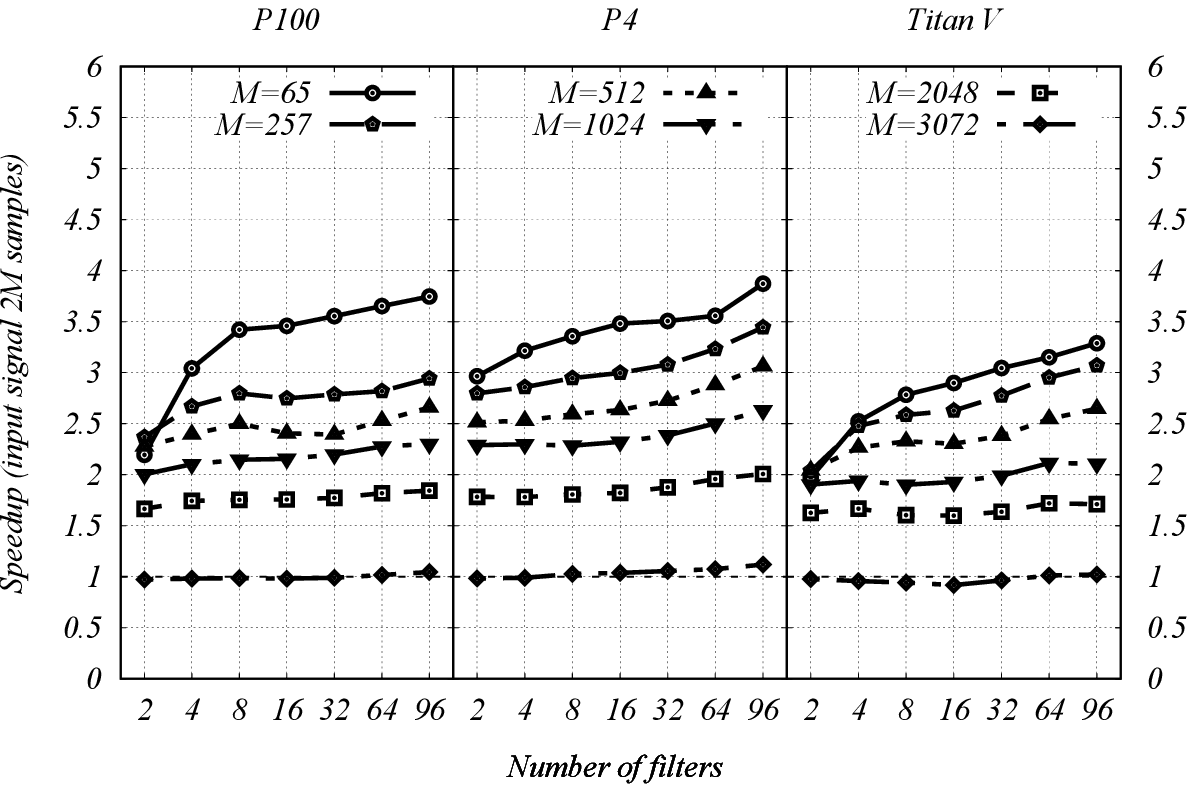}
	    \end{minipage}%
	    \hfill%
	    \begin{minipage}[t]{.485\textwidth}
		    \centering
		    \includegraphics[width=\linewidth]{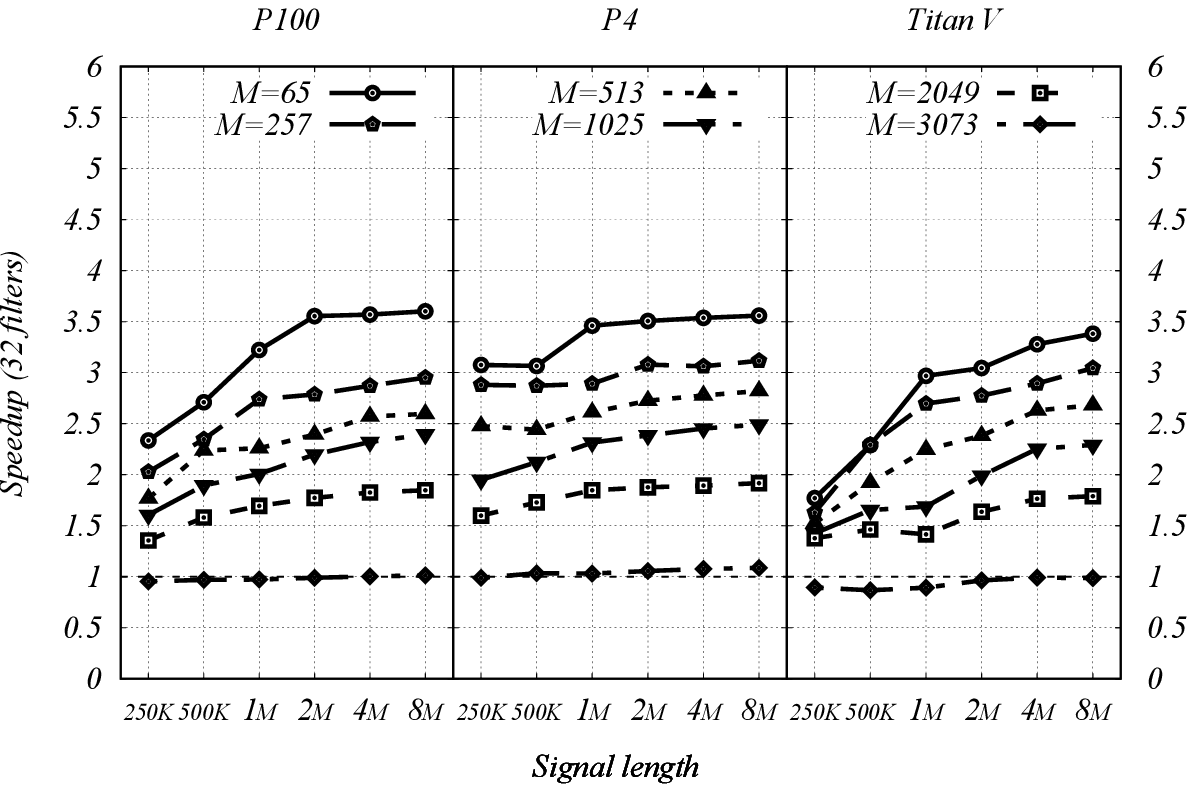}
	    \end{minipage}\\
	    \caption{The speed-up of the R2R SM-OLS convolution with respect to the R2R cuFFT-OLS convolution implementation for different filter lengths vs the number of filters (left), and vs the signal length (right).}
    	\label{fig:R2R_speedups}
    \end{figure}

\begin{figure}[htp]
	    \begin{minipage}[t]{.485\textwidth}
            \centering
            \includegraphics[width=\linewidth]{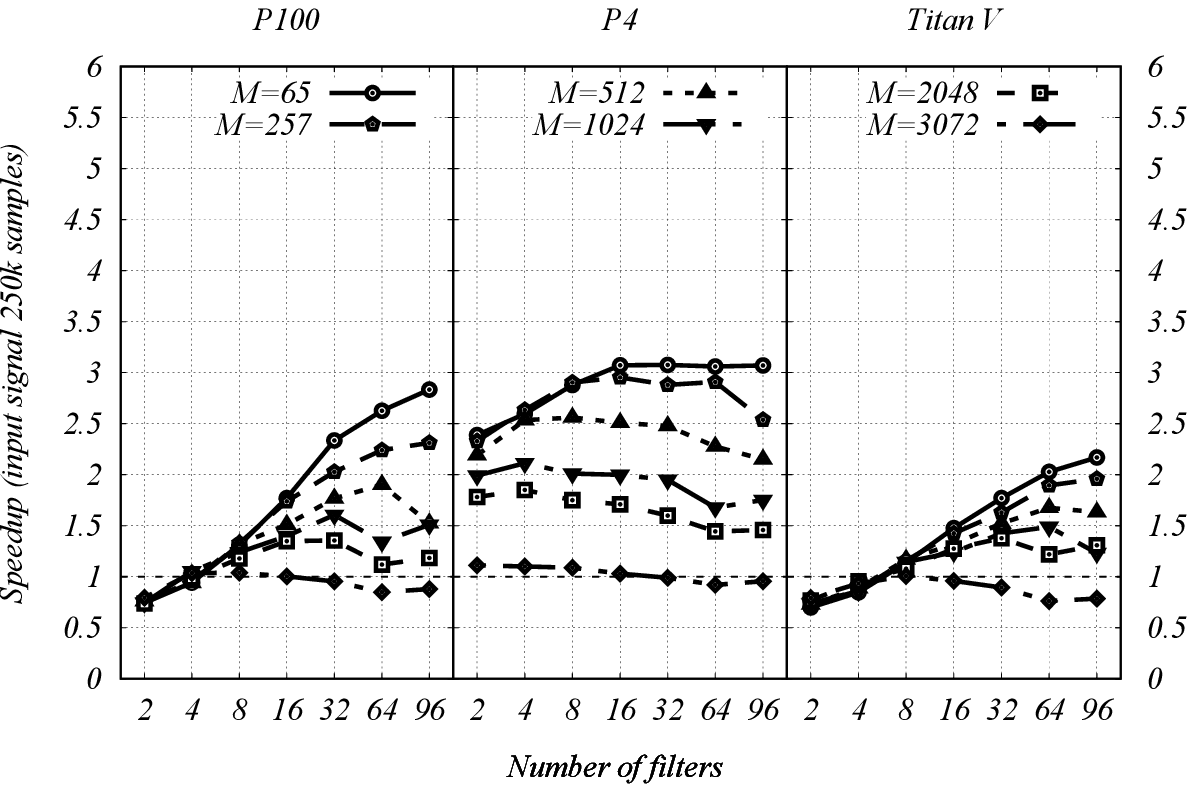}
	    \end{minipage}%
	    \hfill%
	    \begin{minipage}[t]{.485\textwidth}
		    \centering
		    \includegraphics[width=\linewidth]{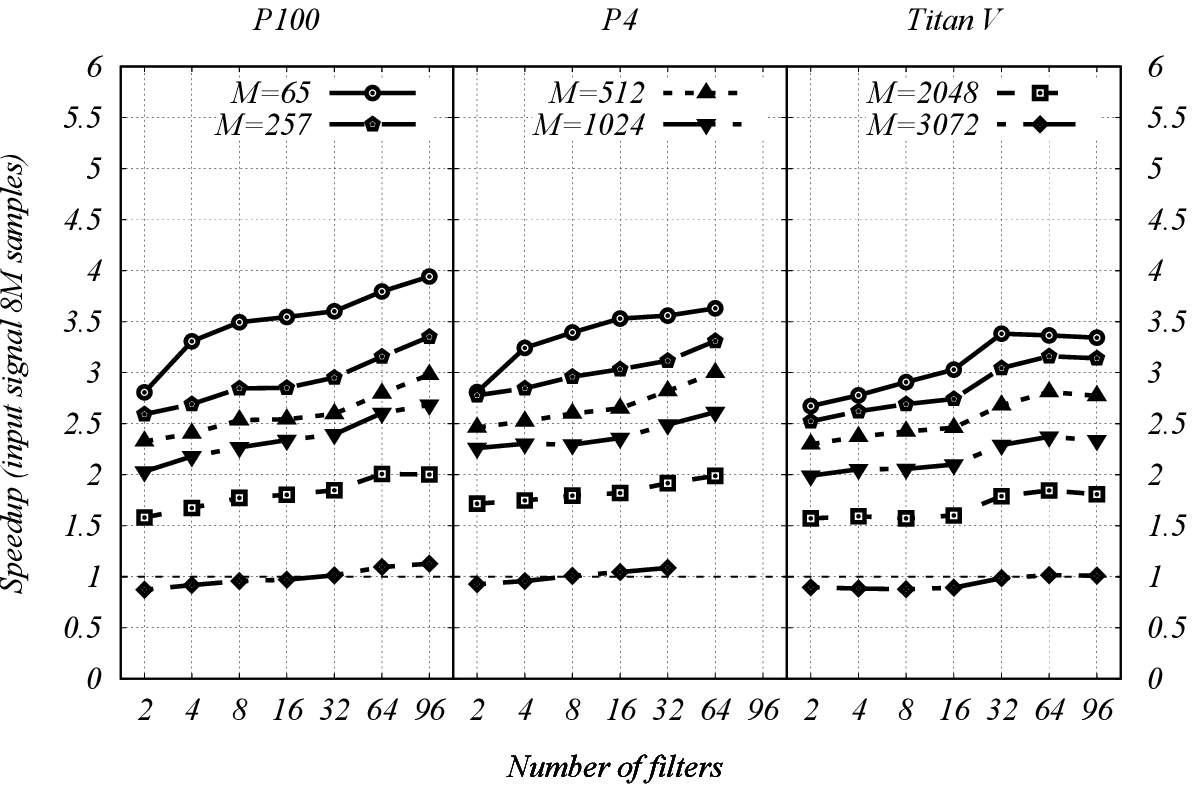}
	    \end{minipage}\\
	    \caption{The speed-up of the R2R SM-OLS convolution with respect to the R2R cuFFT-OLS convolution implementation for different filter lengths vs the number of filters (left) for signal lengths 250k and 8M samples. The number of filters is limited by the amount of device memory the GPU has, this is why there are missing points for P4 GPU (8GB) and TitanV GPU (12GB).}
    	\label{fig:R2R_speedups_other_signallengths}
    \end{figure}

    The speed-up factors for different filter lengths vs the number of filters and vs the input signal length used for 2M signal length are presented in Figure \ref{fig:R2R_speedups}. Speedups for signal lengths other than 2M samples are shown in Figure \ref{fig:R2R_speedups_other_signallengths}.

    \subsection{Non-local Post-processing}
\begin{figure}[htp]
	    \begin{minipage}[t]{.485\textwidth}
            \centering
            \includegraphics[width=\linewidth]{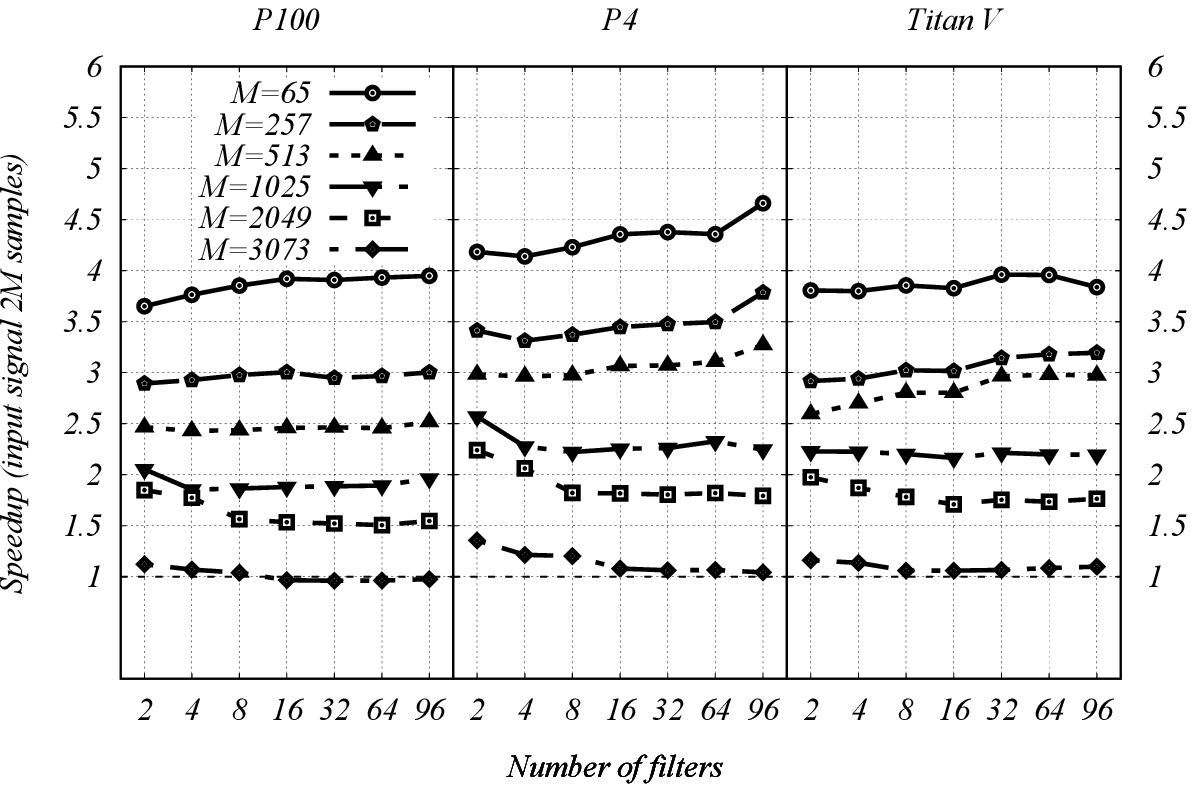}
	    \end{minipage}%
	    \hfill%
	    \begin{minipage}[t]{.485\textwidth}
		    \centering
    		\includegraphics[width=\linewidth]{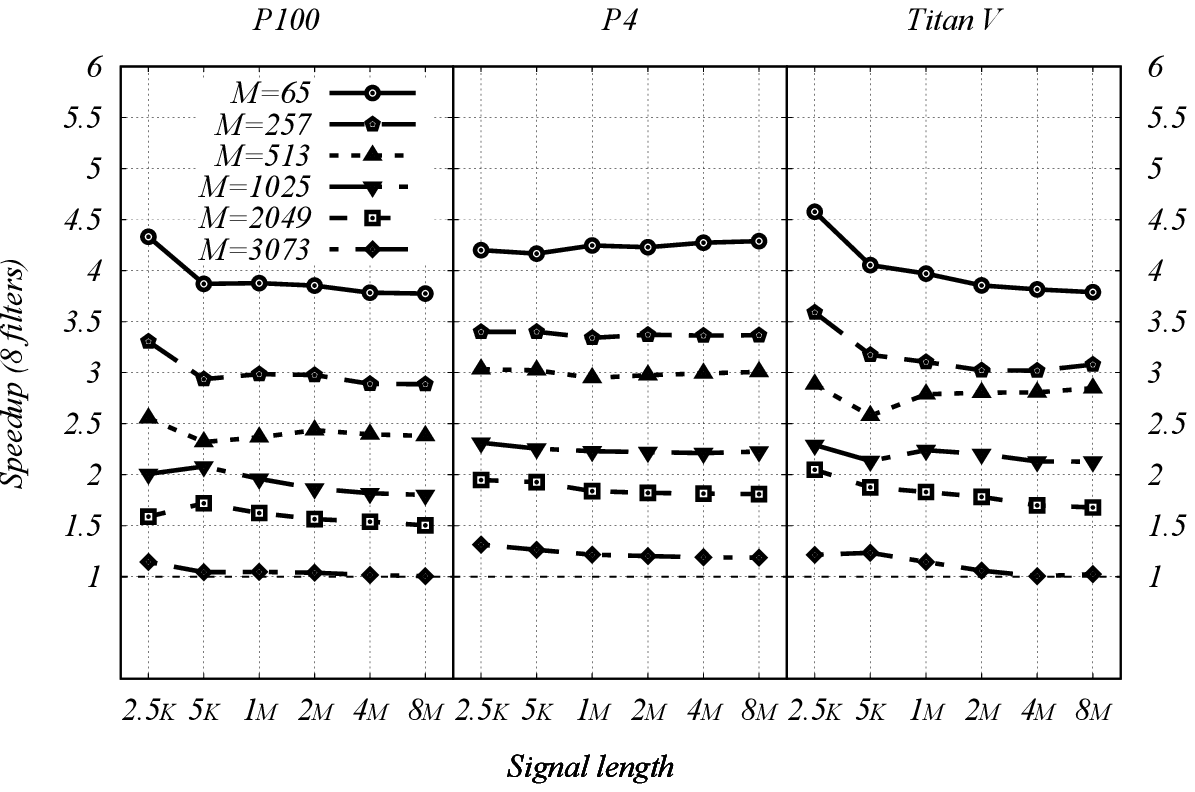}
	    \end{minipage}\\[0.2cm]
	    \begin{minipage}[t]{.485\textwidth}
            \centering
            \includegraphics[width=\linewidth]{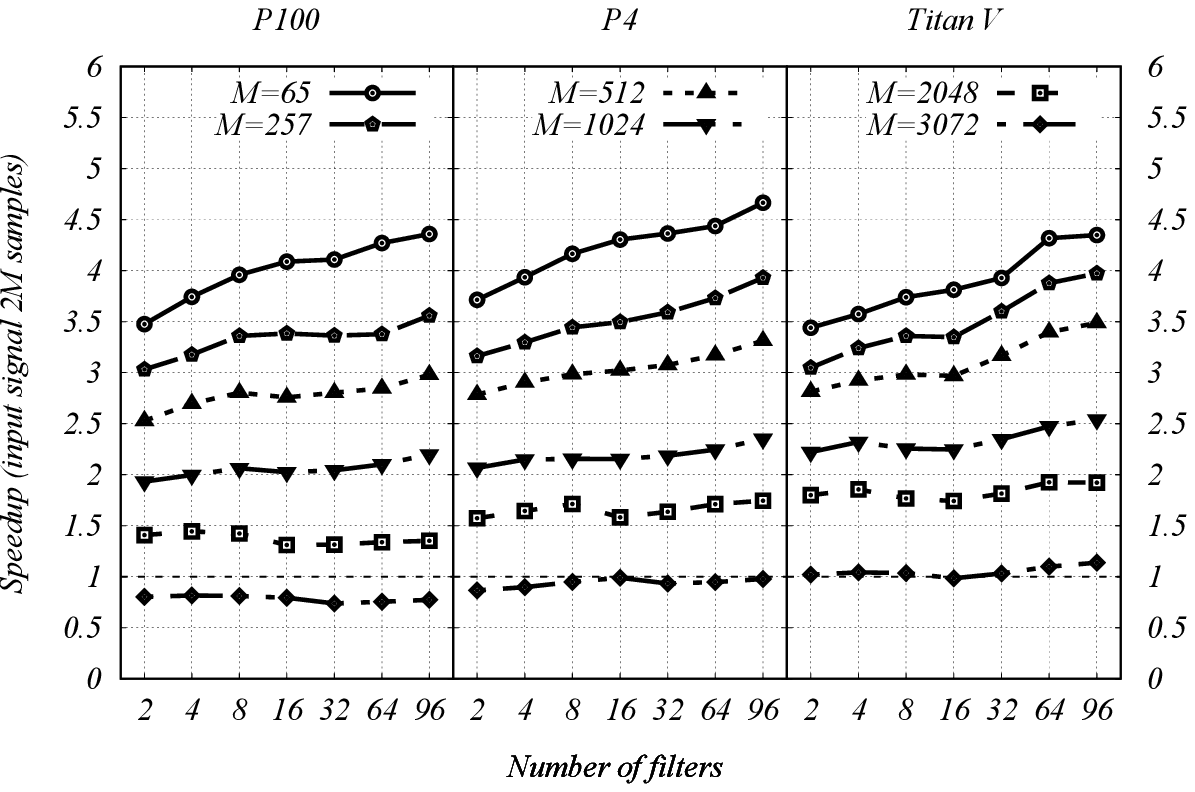}
	    \end{minipage}%
	    \hfill%
	    \begin{minipage}[t]{.485\textwidth}
		    \centering
		    \includegraphics[width=\linewidth]{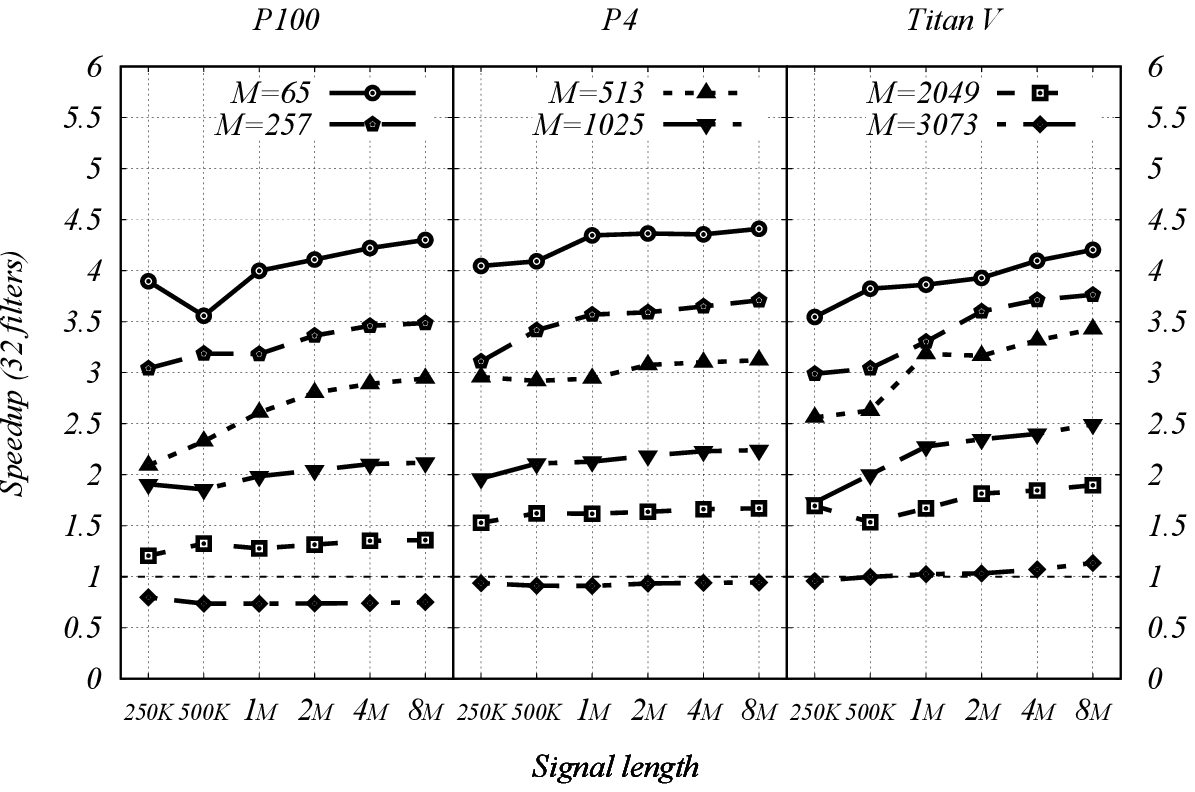}
	    \end{minipage}\\
	    \caption{The speed-up of the SM-OLS over cuFFT-OLS when non-local post-processing is included into consideration. The speed-up for C2C convolution is at the top and speed-up for R2R convolution is at the bottom. }
	    \label{fig:non_local_postprocess}
    \end{figure}

    The advantage of the SM-OLS method is that it has access to all output elements of a given segment. This allows us to perform, in addition to per-element post-processing (for example the calculation of the power spectrum), non-local post-processing as well (for example the numerical derivative or interpolation). The non-local post-processing of output data requires access to the immediate or extended neighborhood of the element to be processed. The cuFFT-OLS method with callbacks offers only limited capabilities when an output element needs to access the values of neighboring elements. The achieved speedups with non-local post-processing are shown in Figure \ref{fig:non_local_postprocess} where we have calculated the derivative of the convolved signal. We have chosen to calculate the derivative because it does not require a larger memory footprint for the output, thus the amount of data which needs to be transferred to and back from device memory remains the same.
    
    \subsection{PCI-e latencies}
    The SM-OLS convolution implementation presented here is most efficient when used as a part of larger signal processing/data reduction pipeline. If run independently the execution time, which includes PCI-e transfer times, would be dominated by the time taken to transfer the output data to the host. Further processing of the output data from the convolution output (such as peak finding or candidate selection) would reduce the amount of output data transferred to the host to a point where the transfer of the output data could be hidden by the computations\footnote{Using, for example, CUDA Streams.}.

\section{Discussion}
    The main source of the speedup for our shared memory OLS implementation for one-dimensional convolution is the elimination of device memory accesses during the convolution step in the OLS method. If every other aspect of the computations in SM-OLS and cuFFT-OLS were equal, the elimination of device memory accesses would result in a constant speedup for all filter lengths, the number of filters or signal length, since the only difference between the two cases would be the per-sample device memory accesses which were not realised. In real calculations, there are many other effects which affect the speedup of our shared memory implementation of OLS convolution.
    
    The primary effect is determined by the segment size $N$, which needs to be set appropriately so that the number of aliased samples which are given by the filter length $M$ is proportionally small compared to the number of uncontaminated output samples contained in the output segment. The segment size in the cuFFT-OLS convolution implementation is not limited to any particular size. This is not true for our implementation of the SM-OLS convolution. Our SM-OLS is limited to a segment length of 4096 samples. This limitation is imposed by the size of the shared memory and also by the number of samples we are able to process per thread. 
    
    If we fix the segment size $N$ then any increase in filter length leads to a decrease in the number of correct output samples per segment, thus more segments are required to calculate the whole convolution. The effect of this can be observed in Figure \ref{fig:C2C_time_conv_length} where different black lines represent the execution time of the SM-OLS implementation with a fixed segment (FFT) size. Figure \ref{fig:C2C_time_conv_length} shows that each segment size is optimal only for a limited range of filter lengths and after that, it is better to switch to a different segment size. Since our implementation of the SM-OLS convolution is limited to segment size $N=4096$ we cannot use a longer segment size when the number of correct samples per segment decreases below a certain limit and at that point cuFFT-OLS becomes the better performing implementation.
    
    The caching of filters is also governed by the size of the segment. The filter length in the frequency domain is equal to the size of the segment $N$ so by increasing segment size we are decreasing the number of filters which can be cached by the GPU's fixed size cache at any instant.

    \subsection{Comparison with cuDNN convolution}
    Figure \ref{fig:cuDNN_OLS_time} shows the comparison of the execution time of our SM-OLS convolution implementation and our implementation of convolution via cuDNN library. The execution time scales linearly with the input signal length shown on the left of Figure \ref{fig:cuDNN_OLS_time}. Different scaling can be seen as the number of filters increase. Our implementation of SM-OLS scales linearly but cuDNN has the same execution time up until the number of filters reaches $32$ at which point it scales linearly. This is due to under-utilization of the GPU resources which is most probably caused by different work distribution which favours more filters.
    
    Figure \ref{fig:cuDNN_OLS_speedup} shows speedup factors of SM-OLS convolution implementation over the cuDNN convolution. The speedup factors for different filter lengths (different line types) vs the signal length (on the left of Figure \ref{fig:cuDNN_OLS_speedup}) shows that both implementations scale at the same rate as the signal length increases. Figure \ref{fig:cuDNN_OLS_speedup} indicates that  the cuDNN library is optimised for small filter lengths since convolutions with smaller filters have lower speedups. The reverse is true when SM-OLS convolution is compared to cuFFT-OLS convolution. The high speedups shown on the right of Figure \ref{fig:cuDNN_OLS_speedup} are due to poor scaling of the cuDNN library for a number of filters below $32$.
    
    \subsection{Comparison with cuFFT convolution}
    The execution time, as shown for C2C convolutions in Figure \ref{fig:C2C_time_flops} and for R2R convolutions in Figure \ref{fig:R2R_time_flops}, scales linearly with an increasing number of filters and increasing input signal length. Both implementations achieve roughly constant performance in the number of processed elements per second past 16 filters or a signal length of two million samples.
    
    The speedup factors of SM-OLS convolution over cuFFT-OLS convolution are shown for C2C convolutions in Figure \ref{fig:C2C_speedups} and in Figure \ref{fig:C2C_speedups_other_signallengths} and for R2R convolutions in Figure \ref{fig:R2R_speedups} and in Figure \ref{fig:R2R_speedups_other_signallengths}. The speedup factors are, in the majority of cases, constant and do not change with the number of filters or the length of the input signal. This is because the segment size is not affected by these parameters, the only difference between the two implementations is the number of device memory accesses performed, or rather not-performed, per sample by the SM-OLS implementation. The total number of processed samples, which includes also the aliased samples, might be different between the two implementations due to different segment sizes used, but the ratio of device memory transfers between these two implementations of the OLS method remains constant and as such the speedup remains constant as well.
    
    There are exceptions to this rule. In the case of complex-to-complex convolutions, we observe (in Figure \ref{fig:C2C_speedups} and on the left of Figure \ref{fig:C2C_speedups_other_signallengths}) that for a small number of filters or short signal lengths we have higher speedups. 
    
    The higher speedup for short signal lengths is due to the slower performance of the cuFFT-OLS convolution which under-utilises GPU resources in this regime. The cuFFT-OLS performs best with longer segment sizes (8192) which, for shorter signal lengths, does not provide enough parallelism for the GPU to utilise. The Titan V GPU which has the most SM\footnote{The SM or streaming multiprocessor is a set of computing cores, the exact number of cores depends on the architecture, which executes threads instruction in parallel.} has the highest speedups while P4 GPU which has the fewest SMs is barely affected.
    
    The high speedups for the small filter numbers are caused by the overhead of creating segments in the cuFFT-OLS implementation. This step is, in the case of SM-OLS, included in the GPU kernel and does not create additional device memory accesses.
    
    The situation is different for R2R convolutions. Speedup factors of SM-OLS convolution over cuFFT-OLS convolution are shown in Figure \ref{fig:R2R_speedups} and in Figure \ref{fig:R2R_speedups_other_signallengths}. We see that for cases with short signal lengths the SM-OLS achieves low or below one speedups. This is caused by the under-utilisation of the GPU resources in our SM-OLS implementation. In the case of R2R convolutions, we are able to convolve a segment of size $N$ with an FFT size $N/2$ \cite{press1992num}, meaning that we are able to fit (depending on the FFT size) up to four thread-blocks per SM which leads to under-utilization even for signal sizes of 500k samples. This can be best observed in Figure \ref{fig:R2R_speedups_other_signallengths} on the left where we show speedups for short signal lengths (250k). GPU cards which are most affected (TitanV GPU, P100 GPU) have also the most SMs, while the P4 GPU with a smaller number of SMs shows speedups comparable to what we can see in Figure \ref{fig:R2R_speedups}. 
    
    Lastly, our SM-OLS has lower performance for shorter filters. This is due to shared memory bank conflicts in our shared memory implementation of the Stockham FFT algorithm. These shared memory bank conflicts occur in the first few iterations of the algorithm. The execution time of these first few iterations dominates the execution time of the shorter FFTs and thus decreases the performance of the whole convolution.
    
    \subsection{Non-local post-processing}
    Figure \ref{fig:non_local_postprocess} shows the speedup of SM-OLS over cuFFT-OLS when performing a non-local post-processing step. Examples, where this might be required, include interpolation of the output or numerical differentiation (which we have used to demonstrate this). The change in the performance depends on the filter size used. The speedup can also decrease when compared to convolution without non-local post-processing. This can be seen for P100 and P4 GPUs when performing real-to-real convolutions with filters longer than 1025 samples, but for shorter filter lengths the speedup can be as great as 30\% as in the case of the Titan V GPU for filter lengths 257 and 513.


\section{Conclusions}
We have presented an implementation of the shared memory overlap-and-save method for the one-dimensional convolution of a large data set with a set of short filters. We have demonstrated a significant speed-up for our shared memory implementation of overlap-and-save, over an implementation of the overlap-and-save method which uses a vendor-supplied FFT library (cuFFT). We have also demonstrated a speedup in the calculation of convolution over a vendor-supplied library for deep neural network primitives (cuDNN) for NVIDIA GPUs. This work has been used to enable real-time data processing in AstroAccelerate software package \cite{AstroAccelerate_2019_2556573} that performs the Fourier Domain Acceleration Search for the Square Kilometre Array \cite{Sofia:2018:FDAS, 2017arXiv171110855A, 2015arXiv151107343D}. Considering the significance of convolution in signal processing this implementation could have a noticeable impact in fields such as natural language processing, monitoring and listening services, speech recognition or pattern matching.

Future work includes the incorporation of the shared memory FFT presented in this paper, into our implementation of a polyphase filter \cite{2016A&C....16....1A} to increase its data throughput. 

\section*{Acknowledgements}
This work has received support from an STFC Grant (ST/R000557/1). The authors would also like to acknowledge the use of the University of Oxford Advanced Research Computing (ARC) \cite{Richards:2015:ARC} facility in carrying out this work. This work is supported by a Leverhulme Trust Project Grant (ARTEMIS: Real-time discovery in Radio Astronomy).


\bibliography{FFT}

\end{document}